\documentclass[preprint,12pt]{elsarticle}

\usepackage{amssymb}
\usepackage{graphicx}
\usepackage{mathrsfs}
\usepackage{amsfonts}
\usepackage{graphicx}
\usepackage{amssymb}
\usepackage{amsmath}
\usepackage{color}
\usepackage[misc]{ifsym}

\newtheorem{open problem}{Open Problem}
\newtheorem{lemma}{Lemma}
\newtheorem{proposition}{Proposition}
\newtheorem{corollary}{Corollary}
\newtheorem{theorem}{Theorem}
\newtheorem{example}{Example}
\newtheorem{remark}{Remark}

\journal{a possible journal}

\begin{document}

\begin{frontmatter}



\title{Constructions of linear codes with two or three weights from vectorial dual-bent functions}


\author{Jiaxin Wang $^{a}$, Zexia Shi $^{b}$, Yadi Wei $^{c}$, Fang-Wei Fu $^{d}$}

\address[a]{Chern Institute of Mathematics and LPMC, Nankai University, Tianjin, 300071, China\\
 email: wjiaxin@mail.nankai.edu.cn}
\address[b]{School of Mathematical Sciences, Hebei Normal University, Shijiazhuang, 050024, China\\
email: shizexia@mail.nankai.edu.cn}
\address[c]{Chern Institute of Mathematics and LPMC, Nankai University, Tianjin, 300071, China\\
 email: wydecho@mail.nankai.edu.cn}
\address[d]{Chern Institute of Mathematics and LPMC, Nankai University, Tianjin, 300071, China\\
 email: fwfu@nankai.edu.cn}

\begin{abstract}
  Linear codes with a few weights are an important class of codes in coding theory and have attracted a lot of attention. In this paper, we present several constructions of $q$-ary linear codes with two or three weights from vectorial dual-bent functions, where $q$ is a power of an odd prime $p$. The weight distributions of the constructed $q$-ary linear codes are completely determined. We illustrate that some known constructions in the literature can be obtained by our constructions. In some special cases, our constructed linear codes can meet the Griesmer bound. Furthermore, based on the constructed $q$-ary linear codes, we obtain secret sharing schemes with interesting access structures.
\end{abstract}


\end{frontmatter}


\section{Introduction}\label{Section1}
Throughout this paper, let $\mathbb{F}_{p}^{r}$ be the vector space of the $r$-tuples over $\mathbb{F}_{p}$, $\mathbb{F}_{p^r}$ be the finite field with $p^r$ elements, $V_{r}$ be an $r$-dimensional vector space over $\mathbb{F}_{p}$, and $\langle \cdot \rangle_{r}$ denote a (non-degenerate) inner product of $V_{r}$, where $p$ is a prime, and  $r$ is a positive integer. In this paper, when $V_{r}=\mathbb{F}_{p}^{r}$, let $\langle a, b\rangle_{r}=a \cdot b=\sum_{i=1}^{r}a_{i}b_{i}$, where $a=(a_{1}, \dots, a_{r}), b=(b_{1}, \dots, b_{r})\in \mathbb{F}_{p}^{r}$; when $V_{r}=\mathbb{F}_{p^r}$, let $\langle a, b\rangle_{r}=Tr_{1}^{r}(ab)$, where $a, b \in \mathbb{F}_{p^r}$, $Tr_{m}^{r}(\cdot)$ denotes the trace function from $\mathbb{F}_{p^r}$ to $\mathbb{F}_{p^m}$, $m$ is a divisor of $r$; when $V_{r}=V_{r_{1}}\times \dots \times V_{r_{t}} (r=\sum_{i=1}^{t}r_{i})$, let $\langle a, b\rangle_{r}=\sum_{i=1}^{t}\langle a_{i}, b_{i}\rangle_{r_{i}}$, where $a=(a_{1}, \dots, a_{t}), b=(b_{1}, \dots, b_{t})\in V_{r}$.

Let $q$ be a power of a prime, and $\mathcal{C}$ be a $q$-ary $[n, k, d]$ linear code, that is, $\mathcal{C}$ is a subspace of $\mathbb{F}_{q}^{n}$ with dimension $k$ and minimum Hamming distance $d$. For any $1\leq i \leq n$, let $A_{i}$ denote the number of codewords in $\mathcal{C}$ whose Hamming weight is $i$. The sequence $(1, A_{1}, \dots, A_{n})$ is called the weight distribution of $\mathcal{C}$, which contains information on the probability of error detection and correction. The code $\mathcal{C}$ is called $t$-weight if $|\{1\leq i\leq n: A_{i}\neq 0\}|=t$.

Linear codes with few weights are an important class of codes in coding theory, which have applications in authentication codes \cite{Ding3}, secret sharing schemes \cite{Anderson,Carlet2,K.Ding,Yuan1}, association schemes \cite{Calderbank1} and strongly regular graphs \cite{Calderbank2}. Due to the importance of linear codes with few weights, they have been extensively studied. Among the methods to construct linear codes with few weights, one of which is based on cryptographic functions such as perfect nonlinear functions over $\mathbb{F}_{p^r}$ (where $p$ is an odd prime) \cite{Carlet2,Feng,Yuan2}, almost perfect nonlinear functions over $\mathbb{F}_{2^r}$ \cite{Carlet1,Ding1}, bent functions from $V_{r}$ to $\mathbb{F}_{p}$ \cite{Ding2,K.Ding,Heng,Mesnager1,Tang,Ozbudak}, and plateaued functions from $V_{r}$ to $\mathbb{F}_{p}$ \cite{Mesnager2,Mesnager3}. For a exhaustive survey on constructing linear codes with few weights from cryptographic functions, we refer to \cite{Li}. In this paper, we present several constructions of $q$-ary linear codes with two or three weights from the so-called vectorial dual-bent functions proposed in \cite{Cesmelioglu2} for which the weight distributions are completely determined, where $q$ is a power of an odd prime $p$. We illustrate that some known constructions in the literature can be obtained by our constructions. In some special cases, our constructed linear codes can meet the Griesmer bound. Furthermore, based on the constructed $q$-ary linear codes, we obtain secret sharing schemes with interesting access structures.

The rest of the paper is organized as follows. In Section 2, we introduce the needed preliminaries. In Section 3.1, we give a construction of $q$-ary linear codes with three-weight. In Section 3.2, we give constructions of $q$-ary linear codes with two-weight. In Section 4, we make a conclusion.

\section{Preliminaries}\label{Section2}
In this section, we introduce the needed prelimilaries. In the sequel, let $q$ be a power of an odd prime $p$, and let $\epsilon=\sqrt{(-1)^\frac{p-1}{2}}$ unless otherwise stated.
\subsection{Vectorial dual-bent functions}\label{Section2.1}
A function $f$ from $V_{r}$ to $V_{m}$ is called a vectorial $p$-ary function, or simply $p$-ary function when $m=1$. The Walsh transform of a $p$-ary function $f: V_{r}\rightarrow \mathbb{F}_{p}$ is defined as
\begin{equation*}
  W_{f}(a)=\sum_{x \in V_{r}}\zeta_{p}^{f(x)-\langle a, x\rangle_{r}}, a \in V_{r},
\end{equation*}
where $\zeta_{p}=e^{\frac{2 \pi \sqrt{-1}}{p}}$ is the complex primitive $p$-th root of unity.

A $p$-ary function $f: V_{r}\rightarrow \mathbb{F}_{p}$ is called bent if $|W_{f}(a)|=p^{\frac{r}{2}}$ for any $a \in V_{r}$. A vectorial $p$-ary function $F: V_{r}\rightarrow V_{m}$ is called vectorial bent if all component functions $F_{c}, c \in V_{m}^{*}=V_{m}\backslash \{0\}$ defined as $F_{c}(x)=\langle c, F(x)\rangle_{m}$ are bent. Note that if $f: V_{r}\rightarrow \mathbb{F}_{p}$ is bent, then so are $cf, c \in \mathbb{F}_{p}^{*}$, that is, bent functions are vectorial bent. The Walsh transform of a $p$-ary bent function $f: V_{r}\rightarrow \mathbb{F}_{p}$ satisfies that for any $a \in V_{r}$,
\begin{equation*}
  W_{f}(a)=\left\{\begin{split}
                     \pm p^{\frac{r}{2}}\zeta_{p}^{f^{*}(a)}, & \ \text{  if } p \equiv 1 \ (mod \ 4) \text{ or } r \text{ is even},\\
                     \pm \sqrt{-1} p^{\frac{r}{2}} \zeta_{p}^{f^{*}(a)}, & \ \text{  if } p \equiv 3 \ (mod \ 4) \text{ and } r  \text{ is odd},
                  \end{split}\right.
\end{equation*}
where $f^{*}$ is a function from $V_{r}$ to $\mathbb{F}_{p}$, called the dual of $f$. A $p$-ary bent function $f: V_{r}\rightarrow \mathbb{F}_{p}$ is said to be weakly regular if $W_{f}(a)=\varepsilon_{f}p^{\frac{r}{2}}\zeta_{p}^{f^{*}(a)}$, where $\varepsilon_{f}$ is a constant independent of $a$, otherwise $f$ is called non-weakly regular. In particular, if $W_{f}(a)=p^{\frac{r}{2}}\zeta_{p}^{f^{*}(a)}$, that is, $\varepsilon_{f}=1$, then $f$ is called regular.

In \cite{Cesmelioglu2}, \c{C}e\c{s}melio\u{g}lu \emph{et al.} introduced vectorial dual-bent functions. A vectorial $p$-ary bent function $F: V_{r}\rightarrow V_{m}$ is said to be vectorial dual-bent if there exists a vectorial bent function $G: V_{r}\rightarrow V_{m}$ such that $(F_{c})^{*}=G_{\sigma(c)}$ for any $c \in V_{m}^{*}$, where $(F_{c})^{*}$ is the dual of the component function $\langle c, F(x)\rangle_{m}$ and $\sigma$ is some permutation over $V_{m}^{*}$. The vectorial bent function $G$ is called a vectorial dual of $F$ and denoted by $F^{*}$. A $p$-ary function $f: V_{r}\rightarrow \mathbb{F}_{p}$ is called an $l$-form if $f(ax)=a^{l}f(x)$ for any $a \in \mathbb{F}_{p}^{*}$ and $x \in V_{r}$, where $1\leq l\leq p-1$ is an integer. In \cite{Cesmelioglu1}, \c{C}e\c{s}melio\u{g}lu \emph{et al.} showed that for a weakly regular $p$-ary bent function $f: V_{r}\rightarrow \mathbb{F}_{p}$ with $f(0)=0$, if $f$ is an $l$-form with $gcd(l-1, p-1)=1$, then $f$ (seen as a vectorial bent function from $V_{r}$ to $V_{1}$) is vectorial dual-bent. Note that a weakly regular bent function $f: V_{r}\rightarrow\mathbb{F}_{p}$ of $l$-form with $f(0)=0$ and $gcd(l-1,p-1)=1$ is just the bent function belonging to $\mathcal{RF}$ in \cite{Tang}.

To the best of our knowledge, there are only a few classes of vectorial dual-bent functions. Below we list the vectorial dual-bent functions given in \cite{Cesmelioglu2,Wang}.
\begin{itemize}
  \item Let $m$ be a divisor of $r'$. Let $F: \mathbb{F}_{p^{r'}} \times \mathbb{F}_{p^{r'}}\rightarrow \mathbb{F}_{p^m}$ be defined as
  \begin{equation}\label{1}
  F(x, y)=Tr_{m}^{r'}(axy^{e}),
  \end{equation}
  where $a \in \mathbb{F}_{p^{r'}}^{*}, gcd(e, p^{r'}-1)=1$. Then $F$ is vectorial dual-bent for which $\varepsilon_{F_{c}}=1$ and $(F_{c})^{*}=(F^{*})_{\sigma(c)}$ for any $c \in \mathbb{F}_{p^m}^{*}$, where $F^{*}(x, y)=Tr_{m}^{r'}(-a^{-u}x^{u}y)$, $\sigma(c)=c^{-u}$ with $eu\equiv 1 \ mod \ (p^{r'}-1)$.
  \item Let $m$ be a divisor of $r'$. Let $F: \mathbb{F}_{p^{r'}} \times \mathbb{F}_{p^{r'}}\rightarrow \mathbb{F}_{p^m}$ be defined as
   \begin{equation}\label{2}
   F(x, y)=Tr_{m}^{r'}(axL(y)),
  \end{equation}
  where $L(x)=\sum a_{i}x^{q^{i}} \ (q=p^m)$ is a $q$-polynomial over $\mathbb{F}_{p^{r'}}$ inducing a permutation of $\mathbb{F}_{p^{r'}}$ and $a \in \mathbb{F}_{p^{r'}}^{*}$. Then $F$ is vectorial dual-bent for which $\varepsilon_{F_{c}}=1$ and $(F_{c})^{*}=(F^{*})_{\sigma(c)}$ for any $c \in \mathbb{F}_{p^m}^{*}$, where $F^{*}(x, y)=Tr_{m}^{r'}(-L^{-1}(a^{-1}x)y)$, $\sigma(c)=c^{-1}$.
  \item In this paper, let $\eta_{r}$ denote the multiplicative quadratic character of $\mathbb{F}_{p^r}$, that is $\eta_{r}(a)=1$ if $a\in \mathbb{F}_{p^r}^{*}$ is a square and $\eta_{r}(a)=-1$ if $a\in \mathbb{F}_{p^r}^{*}$ is a non-square. Let $m$ be a divisor of $r$ and $F: \mathbb{F}_{p^r}\rightarrow \mathbb{F}_{p^m}$ be defined as
   \begin{equation}\label{3}
    F(x)=Tr_{m}^{r}(ax^{2}),
  \end{equation}
  where $a \in \mathbb{F}_{p^r}^{*}$. Then $F$ is vectorial dual-bent for which $\varepsilon_{F_{c}}=(-1)^{r-1}\epsilon^{r}$ $\eta_{r}(ac)$ and $(F_{c})^{*}=(F^{*})_{\sigma(c)}$ for any $c \in \mathbb{F}_{p^m}^{*}$, where $F^{*}(x)=Tr_{m}^{r}(-\frac{x^{2}}{4a})$, $\sigma(c)=c^{-1}$. It is easy to see that when $2m \mid r$, then $\varepsilon_{F_{c}}=-\epsilon^{r}\eta_{r}(a)$ is a constant independent of $c$.
  \item Let $F: \mathbb{F}_{p^m}^{t}\rightarrow \mathbb{F}_{p^m}$ be defined as
  \begin{equation}\label{4}
  F(x_{1}, \dots, x_{t})=a_{1}x_{1}^{2}+\dots+a_{t}x_{t}^{2},
  \end{equation}
  where $a_{i} \in \mathbb{F}_{p^m}^{*}, 1\leq i \leq t$. Then $F$ is vectorial dual-bent for which $\varepsilon_{F_{c}}=(-1)^{(m-1)t}\epsilon^{mt}\eta_{m}(c^{t}a_{1}\cdots a_{t})$ and $(F_{c})^{*}=(F^{*})_{\sigma(c)}$ for any $c \in \mathbb{F}_{p^m}^{*}$, where $F^{*}(x_{1}, \dots, x_{t})=-\frac{x_{1}^{2}}{4a_{1}}-\dots-\frac{x_{t}^{2}}{4a_{t}}$, $\sigma(c)=c^{-1}$. We can see that when $t$ is even, $\varepsilon_{F_{c}}=\epsilon^{mt}\eta_{m}(a_{1}\cdots a_{t})$ is a constant independent of $c$.
  \item Consider a $2r'$-dimensional vector space $\mathbb{F}=\mathbb{F}_{p^{r'}} \times \mathbb{F}_{p^{r'}}$ (or $\mathbb{F}=\mathbb{F}_{p^{2r'}}$) over $\mathbb{F}_{p}$. A partial spread of $\mathbb{F}$ is a collection of subspaces $U_{0}, U_{1}, \dots, U_{w}$ of $\mathbb{F}$ of dimension $r'$ such that $U_{i} \cap U_{j}=\{0\}$ if $i\neq j$. If $w=p^{r'}$, then the collection of subspaces $U_{0}, U_{1}, \dots, U_{p^{r'}}$ is called a spread. For a subspace $U$ of $\mathbb{F}$, let $U^{\perp}=\{z \in \mathbb{F}: \langle z, x \rangle_{2r'}=0 \ \text{for all} \ x \in U\}$. For a positive integer $m\leq r'$, let $g$ be a balanced function from $\{1, 2, \dots, p^{r'}\}$ to $V_{m}$, that is, for any $a \in V_{m}$, $|\{1\leq i\leq p^{r'}: g(i)=a\}|=p^{r'-m}$. Suppose $g(i)=\gamma_{i}, 1\leq i \leq p^{r'}$. Let $\gamma_{0}$ be an arbitrary element of $V_{m}$. Define $F: \mathbb{F}\rightarrow V_{m}$ as
   \begin{equation*}
   F(z)=\gamma_{i} \ \text{if} \ z \in U_{i}, z \neq 0, 1\leq i \leq p^{r'}, \ \text{and} \ F(z)=\gamma_{0} \ \text{if} \ z \in U_{0},
  \end{equation*}
  where $U_{0}, U_{1}, \dots, U_{p^{r'}}$ form a spread of $\mathbb{F}$. Then $F$ is called a vectorial partial spread bent function and it is vectorial dual-bent for which $\varepsilon_{F_{c}}=1$ and $(F_{c})^{*}=(F^{*})_{c}$ for any $c \in V_{m}^{*}$, where $F^{*}(z)=\gamma_{i}$ if $z \in U_{i}^{\perp}, z \neq 0, 1\leq i \leq p^{r'}$, and $F^{*}(z)=\gamma_{0}$ if $z \in U_{0}^{\perp}$. An explicit construction is
  \begin{equation}\label{5}
    F(x_{1}, x_{2})=g(\alpha G(x_{1}x_{2}^{p^{r'}-2})),
  \end{equation}
  where $\alpha \in \mathbb{F}_{p^{r'}}^{*}$, $G$ is a permutation over $\mathbb{F}_{p^{r'}}$ with $G(0)=0$, $g$ is a balanced function from $\mathbb{F}_{p^{r'}}$ to $\mathbb{F}_{p^m}$, $m$ is a divisor of $r'$. A dual of $F$ is $F^{*}(x_{1}, x_{2})=g(\alpha G(-x_{1}^{p^{r'}-2}x_{2}))$.
  \item Let $m, r', r''$ be positive integers with $m \mid r', m \mid r''$. Define $H_{i} \ (i \in \mathbb{F}_{p^m}): \mathbb{F}_{p^{r'}}\rightarrow \mathbb{F}_{p^m}$ as
  $H_{0}(x)=Tr_{m}^{r'}(\alpha_{1}x^{2})$, $H_{i}(x)=Tr_{m}^{r'}(\alpha_{2}x^{2})$ if $i$ is a square in $\mathbb{F}_{p^m}^{*}$, $H_{i}(x)=Tr_{m}^{r'}(\alpha_{3}x^{2})$ if $i$ is a non-square in $\mathbb{F}_{p^m}^{*}$, where $\alpha_{j} \in \mathbb{F}_{p^{r'}}^{*}, 1\leq j\leq 3$. Define $G: \mathbb{F}_{p^{r''}} \times \mathbb{F}_{p^{r''}}\rightarrow \mathbb{F}_{p^m}$ as $G(y_{1}, y_{2})=Tr_{m}^{r''}(\beta y_{1}L(y_{2}))$, where $\beta \in \mathbb{F}_{p^{r''}}^{*}$ and $L(x)=\sum a_{i}x^{q^{i}} \ (q=p^m)$ is a $q$-polynomial over $\mathbb{F}_{p^{r''}}$ inducing a permutation of $\mathbb{F}_{p^{r''}}$. Let $\gamma$ be a nonzero element of $\mathbb{F}_{p^{r''}}$. Then $F: \mathbb{F}_{p^{r'}} \times \mathbb{F}_{p^{r''}} \times \mathbb{F}_{p^{r''}}\rightarrow \mathbb{F}_{p^m}$ defined as
   \begin{equation}\label{6}
   F(x, y_{1}, y_{2})=H_{Tr_{m}^{r''}(\gamma y_{2}^{2})}(x)+G(y_{1}, y_{2})
  \end{equation}
  is a vectorial dual-bent function for which $(F_{c})^{*}=(F^{*})_{\sigma(c)}$ for any $c \in \mathbb{F}_{p^m}^{*}$, where $\sigma(c)=c^{-1}$, $F^{*}(x, y_{1}, y_{2})=R_{Tr_{m}^{r''}(\gamma(L^{-1}(\beta^{-1}y_{1}))^{2})}(x)-Tr_{m}^{r''}(L^{-1}(\beta^{-1}y_{1})y_{2})$, $R_{0}(x)=Tr_{m}^{r'}(-4^{-1}\alpha_{1}^{-1}x^{2})$, $R_{i}(x)=Tr_{m}^{r'}(-4^{-1}\alpha_{2}^{-1}$ $x^{2})$ if $i$ is a square in $\mathbb{F}_{p^m}^{*}$, $R_{i}(x)=Tr_{m}^{r'}(-4^{-1}\alpha_{3}^{-1}x^{2})$ if $i$ is a non-square in $\mathbb{F}_{p^m}^{*}$. When $2m \mid r'$ and $\alpha_{j}, 1\leq j \leq 3$ are all squares or non-squares in $\mathbb{F}_{p^{r'}}^{*}$, then $\varepsilon_{F_{c}}=-\epsilon^{r'}\eta_{r'}(\alpha_{1})$ is a constant independent of $c$.
\end{itemize}

It is not hard to see that the following condition is easily satisfied for vectorial dual-bent functions listed by (1)-(6).

\textbf{Condition A:} Let $p$ be an odd prime, $m, r_{j}, 1\leq j \leq t$ be positive integers and $r\triangleq \sum_{j=1}^{t}r_{j}$ for which $r\geq 4$ is even, $m\leq \frac{r}{2}, m \mid r_{j}, 1\leq j\leq t$, and let $V_{r}=\mathbb{F}_{p^{r_{1}}}\times \mathbb{F}_{p^{r_{2}}}\times \cdots \times \mathbb{F}_{p^{r_{t}}}$. Let $F: V_{r}\rightarrow\mathbb{F}_{p^m}$ be a vectorial dual-bent function for which $F(0)=0, F(-x)=F(x)$, all the component functions of $F$ are regular or weakly regular but not regular (that is, $\varepsilon_{F_{c}}=\varepsilon$ for all $c \in \mathbb{F}_{p^m}^{*}$, where $\varepsilon \in \{\pm1\}$ is a constant independent of $c$), $(F_{c})^{*}=(F^{*})_{\sigma(c)}$ with $\sigma(c)=c^{-e}$, $F^{*}(cx)=c^{e+1}F^{*}(x)$, $c \in \mathbb{F}_{p^m}^{*}, x \in V_{r}$ for some positive integer $e$ with $gcd(e, p^m-1)=1$.

Note that by the proof of Proposition 4 of \cite{Tang}, we have $F^{*}(0)=0$ if $F$ is a vectorial dual-bent function satisfying the Condition A.

\begin{lemma}[\cite{Wang}]\label{Lamma1}
Let $F: V_{r}\rightarrow \mathbb{F}_{p^m}$ and $D_{i}=|\{x \in V_{r}: F(x)=i\}|$, for  $i \in \mathbb{F}_{p^m}$. For any $c \in \mathbb{F}_{p^m}^{*}$, let $F_{c}$ denote the component function of $F$, that is, $F_{c}(x)=Tr_{1}^{m}(cF(x))$. Then
\begin{equation*}
  |D_{i}|=p^{r-m}+p^{-m}\sum_{c \in \mathbb{F}_{p^m}^{*}}W_{F_{c}}(0)\zeta_{p}^{-Tr_{1}^{m}(ci)}, \ i \in \mathbb{F}_{p^m}.
\end{equation*}
If $F$ is a vectorial dual-bent function satisfying the Condition A, then
\begin{equation*}
  |D_{i}|=p^{r-m}+\varepsilon p^{\frac{r}{2}-m}(\delta_{0}(i)p^m-1), i \in \mathbb{F}_{p^m},
\end{equation*}
where $\delta_{0}(0)=1$ and $\delta_{0}(i)=0$ if $i\in \mathbb{F}_{p^m}^{*}$.
\end{lemma}

\subsection{Linear codes}\label{Section2.2}
For a vector $a=(a_{1}, \dots, a_{n})\in \mathbb{F}_{q}^{n}$, let $supp(a)$ denote its support, that is, $supp(a)=\{1\leq i\leq n: a_{i}\neq 0\}$. The Hamming weight of $a$, denoted by $wt(a)$, is the size of its support. For two vectors $a, b \in \mathbb{F}_{q}^{n}$, the Hamming distance $d(a, b)$ between $a$ and $b$ is defined as $d(a, b)=wt(a-b)$. Let $\mathcal{C}$ be a $q$-ary $[n, k]$ linear code. The minimum Hamming distance $d$ of $\mathcal{C}$ is defined as $d=min\{d(a, b): a, b \in \mathcal{C}, a \neq b\}=min\{wt(c): c\in \mathcal{C}, c \neq 0\}$. The dual code of $\mathcal{C}$ is defined by $\mathcal{C}^{\perp}=\{u\in \mathbb{F}_{q}^{n}:u \cdot v=0\ \text{for all} \ u\in \mathcal{C}\}$, where $\cdot$ is the standard inner product on $\mathbb{F}_{q}^{n}$. Note that $\mathcal{C}^{\perp}$ is a $q$-ary $[n, n-k]$ linear code. A codeword $u\in \mathcal{C}$ is said to cover a codeword $v\in \mathcal{C}$ if $supp(v)\subseteq supp(u)$. If a nonzero codeword $u\in \mathcal{C}$ only covers its scalar multiples, then $u$ is called a minimal codeword of $\mathcal{C}$. If all nonzero codewords of $\mathcal{C}$ are minimal, then $\mathcal{C}$ is called minimal. There is a sufficient condition on minimal linear codes.

\begin{lemma}[\cite{Ashikhmin}]\label{Lemma2}
In a $q$-ary $[n, k]$ linear code $\mathcal{C}$, let $w_{min}$ and $w_{max}$ be the minimum and maximum nonzero Hamming weights, respectively. If $\frac{w_{min}}{w_{max}}>\frac{q-1}{q}$, then every nonzero codeword of $\mathcal{C}$ is minimal.
\end{lemma}

We recall two well-known bounds on linear codes (see e.g. \cite{Huffman}).

\begin{proposition}[Griesmer Bound]\label{Proposition1}
Let $\mathcal{C}$ be a $q$-ary $[n, k, d]$ linear code. Then the code $\mathcal{C}$ satisfies the following bound:
\begin{equation*}
  n\geq \sum_{i=0}^{k-1}\left\lceil\frac{d}{q^{i}}\right\rceil,
\end{equation*}
where $\lceil\cdot\rceil$ is the ceiling function.
\end{proposition}

\begin{proposition}[Singleton Bound] \label{Proposition2}
Let $\mathcal{C}$ be a $q$-ary $[n, k, d]$ linear code. Then the code $\mathcal{C}$ satisfies the following bound:
\begin{equation*}
  d\leq n-k+1.
\end{equation*}
\end{proposition}

In the end of this subsection, we recall a well-known result on Gaussian periods.

\begin{lemma}[\cite{Lidl}]\label{Lemma3}
Let $S=\{x^{2}: x \in \mathbb{F}_{p^m}^{*}\}$. Then for any $a \in \mathbb{F}_{p^m}^{*}$
\begin{equation*}
  \sum_{x \in S}\zeta_{p}^{Tr_{1}^{m}(ax)}=\begin{cases}
  \frac{(-1)^{m-1}\epsilon^{m}p^\frac{m}{2}-1}{2}, \text{ if } a \in S,\\
  \frac{-(-1)^{m-1}\epsilon^{m}p^\frac{m}{2}-1}{2}, \text{ if } a \notin S,
  \end{cases}
\end{equation*}
where $\epsilon =1$ if $p\equiv 1 \ (mod \ 4)$ and $\epsilon=\sqrt{-1}$ if $p\equiv 3 \ (mod \ 4)$.
\end{lemma}

\subsection{Secret sharing schemes}\label{Section2.3}
Secret sharing schemes were introduced by Blakley and Shamir independently in 1979 \cite{Blakley,Shamir}, which play an important role in cryptography. Among the approaches to construct secret sharing schemes, one of which is based on linear codes. The following construction was introduced by Massey \cite{Massey}. Let $\mathcal{C}$ be a $q$-ary $[n, k, d]$ linear code, and $G=[g_{0}, g_{1}, \dots, g_{n-1}]$ be a generator matrix of $\mathcal{C}$. In the secret sharing scheme based on $\mathcal{C}$, the secret is an element of $\mathbb{F}_{q}$, and $n-1$ participants $P_{1}, \dots, P_{n-1}$ and a dealer $D$ are involved. In order to compute the shares with respect to a chosen secret $s\in \mathbb{F}_{q}$ by the dealer $D$, the dealer $D$ chooses randomly a vector $u \in \mathbb{F}_{q}^{k}$ such that $s=ug_{0}$. Then the dealer computes the corresponding codeword $\widetilde{s}=(s, s_{1}, \dots, s_{n-1})=uG$ and gives $s_{i}$ as a share to the participant $P_{i}$ for every $1\leq i \leq n-1$. A collection of participants is called an access set if they can recover the secret with their shares. In particular, a collection of participants is called a minimal access set if they can recover the secret with their shares and any of its proper sub-collection cannot recover the secret. The set of all access sets is called the access structure. The following lemma (see Theorem 17 of \cite{Carlet2}) describes the access structure of the secret sharing scheme based on the dual code of a minimal linear code.

\begin{lemma}[\cite{Carlet2}]\label{Lemma4}
Let $\mathcal{C}$ be a $q$-ary $[n, k, d]$ linear code, and let $G=[g_{0}, g_{1}, \dots, $ $g_{n-1}]$ be its generator matrix. We use $d^{\bot}$ to denote the minimum distance of its dual code $\mathcal{C}^{\bot}$. If every nonzero codeword of $\mathcal{C}$ is minimal, then in the secret sharing scheme based on $\mathcal{C}^{\bot}$, there are altogether $q^{k-1}$ minimal access sets.
\begin{itemize}
  \item Case $d^{\bot}=2$:
     For any $1\leq i \leq n-1$, if $g_{i}$ is a multiple of $g_{0}$, then the participant $P_{i}$ must be in every minimal access set. If $g_{i}$ is not a multiple of $g_{0}$, then the participant $P_{i}$ must be in $(q-1)q^{k-2}$ out of $q^{k-1}$ minimal access sets.
  \item Case $d^{\perp}\geq 3$:
     For any fixed $1\leq t \leq min\{k-1, d^{\bot}-2\}$, every set of $t$ participants is involved in $q^{k-1-t}(q-1)^{t}$ out of $q^{k-1}$ minimal access sets.
\end{itemize}
\end{lemma}

\section{Constructions of $q$-ary linear codes with two or three weights}\label{Section3}
In this section, we present several constructions of $q$-ary linear codes with two or three weights from vectorial dual-bent functions.
\subsection{A construction of $q$-ary linear codes with three weights}\label{Section3.1}
In this subsection, we give a construction of $q$-ary linear codes with three weights.
\begin{theorem}\label{Theorem1}
Let $F: V_{r}\rightarrow \mathbb{F}_{p^m}$ be a vectorial dual-bent function satisfying the Condition A. Let $s$ be a divisor of $m$, $q=p^s$. Define \begin{equation}\label{7}
\begin{split}
  &\mathcal{C}_{F}=\{c_{\alpha, \beta}=(Tr_{s}^{m}(\alpha F(x))-\sum_{j=1}^{t}Tr_{s}^{r_{j}}(\beta_{j}x_{j}))_{x=(x_{1}, \dots, x_{t})\in V_{r}^{*}}:\\
  & \ \ \ \ \ \ \ \ \alpha \in \mathbb{F}_{p^m}, \beta=(\beta_{1}, \dots, \beta_{t}) \in V_{r}\}.
\end{split}
\end{equation}
Then $\mathcal{C}_{F}$ is a $q$-ary three-weight linear code with parameters $[p^r-1, \frac{r+m}{s}]$ and the weight distribution is given in Table 1.
\end{theorem}
\begin{table}\label{1}
  \caption{The weight distribution of $\mathcal{C}_{F}$ defined by (7)}
  \begin{tabular}{|c|c|}
    \hline
    Hamming weight $a$ & Multiplicity $A_{a}$ \\
    \hline
    \hline
    $0$ & $1$ \\
    $(p^s-1)p^{r-s}$ & $p^r-1$ \\
    $(p^s-1)p^{r-s}+\varepsilon p^{\frac{r}{2}-s}$ & $(p^m-1)(p^{r-s}-\varepsilon p^{\frac{r}{2}-s})(p^s-1)$ \\
    $(p^s-1)(p^{r-s}-\varepsilon p^{\frac{r}{2}-s})$ & $(p^m-1)(p^{r-s}+\varepsilon p^{\frac{r}{2}-s}(p^s-1))$ \\
    \hline
  \end{tabular}
\end{table}
\textbf{Proof.}
When $\alpha=0, \beta=(0, \dots, 0)$, then $wt(c_{\alpha, \beta})=0$. When $\alpha=0, \beta=(\beta_{1}, \dots, \beta_{t})\neq(0, \dots, 0)$, assume $\beta_{i}\neq 0$ for some $1\leq i \leq t$. By the well-known fact that $Tr_{s}^{r_{i}}(\beta_{i}x)$ is a balanced function from $\mathbb{F}_{p^{r_{i}}}$ to $\mathbb{F}_{p^s}$, we can see that for any $a \in \mathbb{F}_{p^s}$, $|\{x=(x_{1}, \dots, x_{t})\in V_{r}: \sum_{j=1}^{t}Tr_{s}^{r_{j}}(\beta_{j}x_{j})=a\}|=p^{r-s}$. Thus in this case, $wt(c_{\alpha, \beta})=(p^s-1)p^{r-s}$. When $\alpha\neq 0$, $\beta=(\beta_{1}, \dots, \beta_{t})$, since $F(0)=0$, we have $wt(c_{\alpha, \beta})=p^{r}-N_{\alpha, \beta}$, where
\begin{equation*}
  \begin{split}
   N_{\alpha, \beta}&\triangleq|\{x=(x_{1}, \dots, x_{t})\in V_{r}: Tr_{s}^{m}(\alpha F(x))=\sum_{j=1}^{t}Tr_{s}^{r_{j}}(\beta_{j}x_{j})\}|\\
   &=p^{-s}\sum_{c \in \mathbb{F}_{p^s}}\sum_{x=(x_{1}, \dots, x_{t})\in V_{r}}\zeta_{p}^{Tr_{1}^{s}(c(Tr_{s}^{m}(\alpha F(x))-\sum_{j=1}^{t}Tr_{s}^{r_{j}}(\beta_{j}x_{j})))}\\
   &=p^{-s}\sum_{c \in \mathbb{F}_{p^s}^{*}}\sum_{x=(x_{1}, \dots, x_{t})\in V_{r}}\zeta_{p}^{Tr_{1}^{m}(c\alpha F(x))-\sum_{j=1}^{t}Tr_{1}^{r_{j}}(c\beta_{j}x_{j})}+p^{r-s}\\
    &=p^{-s}\sum_{c \in \mathbb{F}_{p^s}^{*}}W_{F_{c \alpha}}(c\beta_{1}, \dots, c\beta_{t})+p^{r-s}\\
   &=\varepsilon p^{\frac{r}{2}-s}\sum_{c \in \mathbb{F}_{p^s}^{*}}\zeta_{p}^{Tr_{1}^{m}(\sigma(c\alpha)F^{*}(c\beta_{1}, \dots, c\beta_{t}))}+p^{r-s}.
  \end{split}
\end{equation*}
Since $\sigma(c)=c^{-e}$, $F^{*}(cx)=c^{e+1}F^{*}(x)$, $c \in \mathbb{F}_{p^m}^{*}, x \in V_{r}$ for some positive integer $e$ with $gcd(e, p^m-1)=1$, we have $Tr_{1}^{m}(\sigma(c\alpha)F^{*}(c\beta_{1}, \dots, c\beta_{t}))=Tr_{1}^{m}(\alpha^{-e}cF^{*}(\beta_{1}, \dots, \beta_{t}))=Tr_{1}^{s}(cTr_{s}^{m}(\alpha^{-e}F^{*}(\beta_{1}, \dots, \beta_{t})))$ for any $c \in \mathbb{F}_{p^s}^{*}$. Thus
\begin{equation}\label{8}
  \begin{split}
  N_{\alpha, \beta}&=\varepsilon p^{\frac{r}{2}-s}\sum_{c \in \mathbb{F}_{p^s}^{*}}\zeta_{p}^{Tr_{1}^{s}(cTr_{s}^{m}(\alpha^{-e}F^{*}(\beta_{1}, \dots, \beta_{t})))}+p^{r-s}\\
  &=\varepsilon p^{\frac{r}{2}-s}(p^{s}\delta_{0}(Tr_{s}^{m}(\alpha^{-e}F^{*}(\beta_{1}, \dots, \beta_{t})))-1)+p^{r-s}.\\
  \end{split}
\end{equation}
By (8), we have
\begin{equation*}
  wt(c_{\alpha, \beta})=\begin{cases}
  (p^{s}-1)(p^{r-s}-\varepsilon p^{\frac{r}{2}-s}), & \text{ if } Tr_{s}^{m}(\alpha^{-e}F^{*}(\beta_{1}, \dots, \beta_{t}))=0,\\
  (p^s-1)p^{r-s}+\varepsilon p^{\frac{r}{2}-s}, & \text{ if } Tr_{s}^{m}(\alpha^{-e}F^{*}(\beta_{1}, \dots, \beta_{t}))\neq 0.\\
  \end{cases}
\end{equation*}
By the above argument, we have $wt(c_{\alpha, \beta})=0$ if and only if $\alpha=0, \beta=(0, \dots, 0)$. Thus the number of codewords is $p^{r+m}$ and the dimension of $\mathcal{C}_{F}$ is $\frac{r+m}{s}$. It is easy to see that $A_{0}=1$, $A_{(p^s-1)p^{r-s}}=|\{(\alpha, \beta)\in \mathbb{F}_{p^m}\times V_{r}: \alpha=0, \beta\neq (0, \dots, 0)\}|=p^{r}-1$. In the following, we compute the values of $A_{(p^{s}-1)(p^{r-s}-\varepsilon p^{\frac{r}{2}-s})}$ and $A_{(p^s-1)p^{r-s}+\varepsilon p^{\frac{r}{2}-s}}$. For any fixed $\alpha \neq 0$, $Tr_{s}^{m}(\alpha^{-e}F^{*}(\beta))=0$ if and only if $F^{*}(\beta)=0, \alpha^{e}z_{1}, \dots, $ $ \alpha^{e}z_{p^{m-s}-1}$, where $z_{1}, \dots, z_{p^{m-s}-1}$ are the nonzero solutions of $Tr_{s}^{m}(z)=0$. By Lemma 1, for any $a \in \mathbb{F}_{p^m}$ we have
\begin{equation}\label{9}
\begin{split}
  |\{x \in V_{r}: F^{*}(x)=a\}|&=p^{r-m}+p^{-m}\sum_{\alpha \in \mathbb{F}_{p^m}^{*}}W_{(F^{*})_{\alpha}}(0)\zeta_{p}^{-Tr_{1}^{m}(a\alpha)}\\
  &=p^{r-m}+p^{-m}\sum_{\alpha \in \mathbb{F}_{p^m}^{*}}W_{(F_{\sigma^{-1}(\alpha)})^{*}}(0)\zeta_{p}^{-Tr_{1}^{m}(a\alpha)}\\
  &=p^{r-m}+\varepsilon p^{\frac{r}{2}-m}\sum_{\alpha \in \mathbb{F}_{p^m}^{*}}\zeta_{p}^{F_{\sigma^{-1}(\alpha)}(0)-Tr_{1}^{m}(a\alpha)}\\
  &=p^{r-m}+\varepsilon p^{\frac{r}{2}-m}(\delta_{0}(a)p^m-1),\\
\end{split}
\end{equation}
where the second equation is obtained by $(F_{\alpha})^{*}=(F^{*})_{\sigma(\alpha)}$ and in the third equation we use the fact that if $f$ is a weakly regular bent function with dual $f^{*}$, then $f^{*}$ is also a weakly regular bent function with $(f^{*})^{*}(x)=f(-x)$. Therefore, by (9) we have
\begin{equation*}
\begin{split}
  &A_{(p^{s}-1)(p^{r-s}-\varepsilon p^{\frac{r}{2}-s})}\\
  &=|\{(\alpha, \beta)\in \mathbb{F}_{p^m}^{*}\times V_{r}: Tr_{s}^{m}(\alpha^{-e}F^{*}(\beta))=0\}|\\
  &=(p^m-1)(p^{r-m}+\varepsilon p^{\frac{r}{2}-m}(p^m-1)+(p^{r-m}-\varepsilon p^{\frac{r}{2}-m})(p^{m-s}-1))\\
  &=(p^m-1)(p^{r-s}+\varepsilon p^{\frac{r}{2}-s}(p^s-1)),\\
\end{split}
\end{equation*}
and
\begin{equation*}
\begin{split}
  A_{(p^s-1)p^{r-s}+\varepsilon p^{\frac{r}{2}-s}}
  &=|\{(\alpha, \beta) \in \mathbb{F}_{p^m}^{*} \times V_{r}: Tr_{s}^{m}(\alpha^{-e}F^{*}(\beta))\neq 0\}|\\
  &=(p^{m}-1)p^{r}-A_{(p^{s}-1)(p^{r-s}-\varepsilon p^{\frac{r}{2}-s})}\\
  &=(p^m-1)(p^{r-s}-\varepsilon p^{\frac{r}{2}-s})(p^s-1). \ \ \ \Box\\
\end{split}
\end{equation*}

Based on vectorial dual-bent functions listed by (1)-(6), we list the corresponding three-weight linear codes from Theorem 1.
\begin{itemize}
  \item Let $F: \mathbb{F}_{p^{r'}} \times \mathbb{F}_{p^{r'}}\rightarrow \mathbb{F}_{p^m}$ be defined as (1) (respectively, (2), (5)). Then $F$ is vectorial dual-bent satisfying the Condition A and the corresponding value $\varepsilon=1$. Let $s$ be a divisor of $m$, then the corresponding linear code $\mathcal{C}_{F}$ defined by (7) is
  \begin{equation*}
  \begin{split}
  &\mathcal{C}_{F}=\{c_{\alpha, \beta}=(Tr_{s}^{r'}(\alpha ax_{1}x_{2}^{e}-\beta_{1}x_{1}-\beta_{2}x_{2}))_{x=(x_{1}, x_{2})\in (\mathbb{F}_{p^{r'}}\times\mathbb{F}_{p^{r'}})^{*}}:\\
  & \ \ \ \ \ \ \ \ \ \alpha \in \mathbb{F}_{p^m}, \beta=(\beta_{1}, \beta_{2}) \in \mathbb{F}_{p^{r'}}\times\mathbb{F}_{p^{r'}}\},
  \end{split}
  \end{equation*}
  (respectively,
  \begin{equation*}
  \begin{split}
  &\mathcal{C}_{F}=\{c_{\alpha, \beta}=(Tr_{s}^{r'}(\alpha ax_{1}L(x_{2})-\beta_{1}x_{1}-\beta_{2}x_{2}))_{x=(x_{1}, x_{2})\in (\mathbb{F}_{p^{r'}}\times\mathbb{F}_{p^{r'}})^{*}}:\\
  & \ \ \ \ \ \ \ \ \ \alpha \in \mathbb{F}_{p^m}, \beta=(\beta_{1}, \beta_{2}) \in \mathbb{F}_{p^{r'}}\times\mathbb{F}_{p^{r'}}\},\\
  &\mathcal{C}_{F}=\\
  &\{c_{\alpha, \beta}=(Tr_{s}^{m}(\alpha g(\gamma G(x_{1}x_{2}^{p^{r'}-2})))-Tr_{s}^{r'}(\beta_{1}x_{1}+\beta_{2}x_{2}))_{x=(x_{1}, x_{2})\in (\mathbb{F}_{p^{r'}}\times \mathbb{F}_{p^{r'}})^{*}} \\
  & \ \ \ \ \ \ \ \ \ : \alpha \in \mathbb{F}_{p^m}, \beta=(\beta_{1}, \beta_{2}) \in \mathbb{F}_{p^{r'}}\times \mathbb{F}_{p^{r'}}\}
  \end{split}
  \end{equation*}),
  which is a three-weight $p^s$-ary $[p^{2r'}-1, \frac{2r'+m}{s}]$ linear code.
  \item Let $F: \mathbb{F}_{p^{r}}\rightarrow \mathbb{F}_{p^m}$ be defined as (3) with $2m \mid r$. Then $F$ is vectorial dual-bent satisfying the Condition A and the corresponding value $\varepsilon=-\epsilon^{{r}}\eta_{r}(a)$. Let $s$ be a divisor of $m$, then the corresponding linear code $\mathcal{C}_{F}$ defined by (7) is
  \begin{equation}\label{10}
  \begin{split}
  &\mathcal{C}_{F}=\{c_{\alpha, \beta}=(Tr_{s}^{r}(\alpha a x^{2}-\beta x))_{x\in \mathbb{F}_{p^{r}}^{*}}: \alpha \in \mathbb{F}_{p^m}, \beta \in \mathbb{F}_{p^{r}}\},
  \end{split}
  \end{equation}
  which is a three-weight $p^s$-ary $[p^{r}-1, \frac{r+m}{s}]$ linear code. It is easy to see that $\mathcal{C}_{F}$ is a subcode of the linear code $\mathcal{C}$ defined by
  \begin{equation*}
   \mathcal{C}=\{c_{\alpha, \beta}=(Tr_{s}^{r}(\alpha a x^{2}-\beta x))_{x \in \mathbb{F}_{p^{r}}^{*}}: \alpha, \beta \in \mathbb{F}_{p^{r}}\},
  \end{equation*}
  which is defined by (6) of \cite{Carlet2} with $\emph{II}(x)=x^{2}$. To the best of our knowledge, the weight distribution of $\mathcal{C}$ is not yet determined except for $s=1$. Our above result shows that when $s \mid m$ and $2m \mid r$, there is a subcode of $\mathcal{C}$ defined by (10) for which the weight distribution is completely determined.
  \item Let $F: \mathbb{F}_{p^m}^{t}\rightarrow \mathbb{F}_{p^m}$ be defined as (4) with $t$ even. Then $F$ is vectorial dual-bent satisfying the Condition A and the corresponding value $\varepsilon=\epsilon^{mt}\eta_{m}(a_{1}\cdots a_{t})$. Let $s$ be a divisor of $m$, then the corresponding linear code $\mathcal{C}_{F}$ defined by (7) is
  \begin{equation*}
  \begin{split}
  &\mathcal{C}_{F}=\{c_{\alpha, \beta}=(\sum_{i=1}^{t}Tr_{s}^{m}(\alpha a_{i} x_{i}^{2}-\beta_{i} x_{i}))_{x=(x_{1}, \dots, x_{t})\in (\mathbb{F}_{p^m}^{t})^{*}}: \\
  & \ \ \ \ \ \ \ \ \ \alpha \in \mathbb{F}_{p^m}, \beta=(\beta_{1}, \dots, \beta_{t}) \in \mathbb{F}_{p^m}^{t}\},
  \end{split}
  \end{equation*}
  which is a three-weight $p^s$-ary $[p^{mt}-1, \frac{m(t+1)}{s}]$ linear code.
  \item Let $F: \mathbb{F}_{p^{r'}} \times \mathbb{F}_{p^{r''}} \times \mathbb{F}_{p^{r''}}\rightarrow \mathbb{F}_{p^m}$ be defined as (6) for which $2m \mid r'$ and $\alpha_{j}, 1\leq j \leq 3$ are all squares or non-squares in $\mathbb{F}_{p^{r'}}^{*}$. Then $F$ is vectorial dual-bent satisfying the Condition A and the corresponding value $\varepsilon=-\epsilon^{r'}\eta_{r'}(\alpha_{1})$. Let $s$ be a divisor $m$, then the corresponding linear code $\mathcal{C}_{F}$ defined by (7) is
  \begin{small}
  \begin{equation*}
  \begin{split}
  &\mathcal{C}_{F}=\{c_{\alpha, \beta}=\\
  &(Tr_{s}^{m}(\alpha F(x_{1}, x_{2}, x_{3}))-Tr_{s}^{r'}(\beta_{1}x_{1})-Tr_{s}^{r''}(\beta_{2}x_{2}+\beta_{3}x_{3}))_{x=(x_{1}, x_{2}, x_{3})\in (\mathbb{F}_{p^{r'}}\times \mathbb{F}_{p^{r''}}\times \mathbb{F}_{p^{r''}})^{*}} \\
  & : \alpha \in \mathbb{F}_{p^m}, \beta=(\beta_{1}, \beta_{2}, \beta_{3}) \in \mathbb{F}_{p^{r'}}\times \mathbb{F}_{p^{r''}}\times \mathbb{F}_{p^{r''}}\},
  \end{split}
  \end{equation*}
  \end{small}which is a three-weight $p^s$-ary $[p^{r'+2r''}-1, \frac{m+r'+2r''}{s}]$ linear code.
\end{itemize}

We illustrate that $\mathcal{C}_{F}$ defined by (10) can be used to construct optimal linear codes in a special case. When $s=m=\frac{r}{2}$ and $a \in \mathbb{F}_{p^{r}}^{*}$ with $\eta_{r}(a)=\epsilon^{r}$, then the corresponding value $\varepsilon=-1$ and it is easy to verify that the linear code $\mathcal{C}_{F}$ defined by (10) is a three-weight $p^{\frac{r}{2}}$-ary linear code with parameters $[p^{r}-1, 3, (p^{\frac{r}{2}}-1)p^{\frac{r}{2}}-1]$ and meets the Griesmer bound. We give an example.

\begin{example}\label{Example1}
Let $p=3, s=m=2, r=4$, and $F: \mathbb{F}_{3^4}\rightarrow \mathbb{F}_{3^2}$ be defined as $F(x)=Tr_{2}^{4}(x^{2})$, then $F$ is vectorial dual-bent which satisfies the Condition A and the corresponding linear code $\mathcal{C}_{F}$ defined by (7) is
\begin{equation*}
  \mathcal{C}_{F}=\{c_{\alpha, \beta}=(\alpha Tr_{2}^{4}(x^{2})-Tr_{2}^{4}(\beta x))_{x \in \mathbb{F}_{3^4}^{*}}: \alpha \in \mathbb{F}_{3^2}, \beta \in \mathbb{F}_{3^4}\},
\end{equation*}
which is a three-weight $9$-ary linear code with parameters $[80, 3, 71]$ and the weight enumerator is $1+640z^{71}+80z^{72}+8z^{80}$. It is easy to verify that $\mathcal{C}_{F}$ meets the Griesmer bound.
\end{example}

Based on Lemmas 2 and 4, we have the following proposition.

\begin{proposition}\label{Proposition3}
Let $\mathcal{C}_{F}$ be the $p^s$-ary linear code constructed in Theorem 1 with $s\leq \frac{r}{2}-1$, and let $G=[g_{0}, g_{1}, \dots, g_{p^r-2}]$ be its generator matrix. In the secret sharing scheme based on $\mathcal{C}_{F}^{\bot}$, there are altogether $p^{r+m-s}$ minimal access sets. For any $1\leq i \leq p^r-2$, if $g_{i}$ is a multiple of $g_{0}$, then the participant $P_{i}$ must be in every minimal access set. If $g_{i}$ is not a multiple of $g_{0}$, then the participant $P_{i}$ must be in $(p^s-1)p^{r+m-2s}$ out of $p^{r+m-s}$ minimal access sets.
\end{proposition}
\textbf{Proof.}
From Lemma 2, it is easy to verify that the constructed $p^s$-ary linear code of Theorem 1 is minimal if $s\leq \frac{r}{2}-1$. It is easy to see that the minimum distance of $\mathcal{C}_{F}^{\bot}$ is not $1$. By Lemma 1, there exists $x=(x_{1}, \dots, x_{t})$ such that $F(x)=0$. Since $F(-x)=F(x)$, we have $F(-x)=0$. Then $Tr_{s}^{m}(\alpha F(x))-\sum_{j=1}^{t}Tr_{s}^{r_{j}}(\beta_{j}x_{j})+Tr_{s}^{m}(\alpha F(-x))-\sum_{j=1}^{t}Tr_{s}^{r_{j}}(-\beta_{j}x_{j})=0$ for any $\alpha \in \mathbb{F}_{p^m}$, $\beta \in V_{r}$, which implies that the minimum distance of $\mathcal{C}_{F}^{\bot}$ is $2$. Then the conclusion follows from Lemma 4. \ $\Box$

\subsection{Constructions of $q$-ary linear codes with two weights}\label{Section3.2}
In this subsection, we present constructions of $q$-ary linear codes with two weights.
\begin{theorem}\label{Theorem2}
Let $F: V_{r} \rightarrow \mathbb{F}_{p^m}$ be a vectorial dual-bent function satisfying the Condition A. Let $s_{1}, s_{2}$ be positive integers with $s_{2} \mid m, s_{1} \mid s_{2}$, and $s_{2}\neq \frac{r}{2}$ if $\varepsilon=-1$. Let $q=p^{s_{1}}$. Define
\begin{equation}\label{11}
  \mathcal{C}_{D(F)}=\{c_{\alpha}=(\sum_{j=1}^{t}Tr_{s_{1}}^{r_{j}}(\alpha_{j}d_{j}))_{d=(d_{1}, \dots, d_{t}) \in D(F)}: \alpha=(\alpha_{1}, \dots, \alpha_{t})\in V_{r}\},
\end{equation}
where $D(F)=\{x \in V_{r}^{*}: Tr_{s_{2}}^{m}(\lambda F(x))=0\}$, $\lambda \in \mathbb{F}_{p^m}^{*}$. Then $\mathcal{C}_{D(F)}$ defined by (11) is a two-weight $q$-ary linear code with parameters $[p^{r-s_{2}}+\varepsilon p^{\frac{r}{2}-s_{2}}(p^{s_{2}}-1)-1, \frac{r}{s_{1}}]$ and the weight distribution is given in Table 2.
\end{theorem}
\begin{table}\label{2}
  \caption{The weight distribution of $\mathcal{C}_{D(F)}$ defined by (11)}
  \begin{tabular}{|c|c|}
    \hline
    Hamming weight $a$ & Multiplicity $A_{a}$ \\
    \hline
    \hline
    $0$ & $1$ \\
    $(p^{s_{1}}-1)p^{r-s_{1}-s_{2}}$ & $p^{r-s_{2}}+\varepsilon p^{\frac{r}{2}-s_{2}}(p^{s_{2}}-1)-1$ \\
    $(p^{s_{1}}-1)(p^{r-s_{1}-s_{2}}+\varepsilon p^{\frac{r}{2}-s_{1}})$ & $(p^{s_{2}}-1)(p^{r-s_{2}}-\varepsilon p^{\frac{r}{2}-s_{2}})$ \\
    \hline
  \end{tabular}
\end{table}
\textbf{Proof.} Note that $Tr_{s_{2}}^{m}(\lambda F(x))=0$ if and only if $F(x)=0, \lambda^{-1}z_{1}, \dots, $ $\lambda^{-1}z_{p^{m-s_{2}}-1}$, where $z_{1}, \dots, z_{p^{m-s_{2}}-1}$ are the nonzero solutions of $Tr_{s_{2}}^{m}(z)=0$. By Lemma 1, for any $a \in \mathbb{F}_{p^m}$ we have $|\{x \in V_{r}: F(x)=a\}|=p^{r-m}+\varepsilon p^{\frac{r}{2}-m}(\delta_{0}(a)p^m-1)$. Therefore, $|\{x \in V_{r}: Tr_{s_{2}}^{m}(\lambda F(x))=0\}|=
p^{r-m}+\varepsilon p^{\frac{r}{2}-m}(p^m-1)+(p^{m-s_{2}}-1)(p^{r-m}-\varepsilon p^{\frac{r}{2}-m})=p^{r-s_{2}}+\varepsilon p^{\frac{r}{2}-s_{2}}(p^{s_{2}}-1)$ and the length of linear code $\mathcal{C}_{D(F)}$ is $|D(F)|=|\{x \in V_{r}: Tr_{s_{2}}^{m}(\lambda F(x))=0\}|-1=p^{r-s_{2}}+\varepsilon p^{\frac{r}{2}-s_{2}}(p^{s_{2}}-1)-1$. When $\alpha=(0, \dots, 0)$, then $wt(c_{\alpha})=0$. When $\alpha=(\alpha_{1}, \dots, \alpha_{t}) \neq (0, \dots, 0)$, we have $wt(c_{\alpha})=|\{x \in V_{r}: Tr_{s_{2}}^{m}(\lambda F(x))=0\}|-N_{\alpha}$, where
\begin{equation*}
\begin{split}
  N_{\alpha}&\triangleq |\{x=(x_{1}, \dots, x_{t}) \in V_{r}: Tr_{s_{2}}^{m}(\lambda F(x))=0, \sum_{j=1}^{t}Tr_{s_{1}}^{r_{j}}(\alpha_{j}x_{j})=0\}|\\
  &=p^{-s_{1}-s_{2}}\sum_{x=(x_{1}, \dots, x_{t}) \in V_{r}}\sum_{y \in \mathbb{F}_{p^{s_{2}}}}\zeta_{p}^{Tr_{1}^{s_{2}}(yTr_{s_{2}}^{m}(\lambda F(x)))}\\
  & \ \ \ \times\sum_{z \in \mathbb{F}_{p^{s_{1}}}}\zeta_{p}^{Tr_{1}^{s_{1}}(z(-\sum_{j=1}^{t}Tr_{s_{1}}^{r_{j}}(\alpha_{j}x_{j})))}\\
  &=p^{-s_{1}-s_{2}}\sum_{z \in \mathbb{F}_{p^{s_{1}}}}\sum_{y \in \mathbb{F}_{p^{s_{2}}}^{*}}\sum_{x=(x_{1}, \dots, x_{t}) \in V_{r}}\zeta_{p}^{Tr_{1}^{m}(y \lambda F(x))-\sum_{j=1}^{t}Tr_{1}^{r_{j}}(z\alpha_{j}x_{j})}\\
\end{split}
\end{equation*}
\begin{equation*}
\begin{split}
&\ \ +p^{-s_{1}-s_{2}}\sum_{x=(x_{1}, \dots, x_{t}) \in V_{r}}\sum_{z \in \mathbb{F}_{p^{s_{1}}}}\zeta_{p}^{Tr_{1}^{s_{1}}(z(-\sum_{j=1}^{t}Tr_{s_{1}}^{r_{j}}(\alpha_{j}x_{j})))}\\
 &=p^{-s_{1}-s_{2}}\sum_{z \in \mathbb{F}_{p^{s_{1}}}}\sum_{y \in \mathbb{F}_{p^{s_{2}}}^{*}}W_{F_{y\lambda}}(z\alpha)+p^{-s_{1}-s_{2}}\cdot p^{s_{1}}\cdot p^{r-s_{1}}\\
 &=\varepsilon p^{\frac{r}{2}-s_{1}-s_{2}}\sum_{z \in \mathbb{F}_{p^{s_{1}}}}\sum_{y \in \mathbb{F}_{p^{s_{2}}}^{*}}\zeta_{p}^{Tr_{1}^{m}(\sigma(y\lambda)F^{*}(z\alpha))}+p^{r-s_{1}-s_{2}}\\
 &=\varepsilon p^{\frac{r}{2}-s_{1}-s_{2}}\sum_{z \in \mathbb{F}_{p^{s_{1}}}^{*}}\sum_{y \in \mathbb{F}_{p^{s_{2}}}^{*}}\zeta_{p}^{Tr_{1}^{m}(\sigma(y\lambda)F^{*}(z\alpha))}+\varepsilon p^{\frac{r}{2}-s_{1}-s_{2}}(p^{s_{2}}-1)\\
 & \ \ +p^{r-s_{1}-s_{2}}\\
 &=\varepsilon p^{\frac{r}{2}-s_{1}-s_{2}}\sum_{z \in \mathbb{F}_{p^{s_{1}}}^{*}}\sum_{y \in \mathbb{F}_{p^{s_{2}}}^{*}}\zeta_{p}^{Tr_{1}^{m}(\lambda^{-e}(\frac{z}{y})^{e}zF^{*}(\alpha))}+\varepsilon p^{\frac{r}{2}-s_{1}-s_{2}}(p^{s_{2}}-1)\\
  & \ \ +p^{r-s_{1}-s_{2}},
\end{split}
\end{equation*}
where the last equation is obtained by $\sigma(c)=c^{-e}, F^{*}(cx)=c^{e+1}F^{*}(x), c \in \mathbb{F}_{p^m}^{*}, x \in V_{r}$ with $gcd(e, p^m-1)=1$. For any fixed $z \in \mathbb{F}_{p^{s_{1}}}^{*}$, since $s_{1} \mid s_{2}$ and $s_{2}\mid m$ and $gcd(e, p^m-1)=1$, we have that $P(y)=(\frac{z}{y})^{e}$ is a permutation over $\mathbb{F}_{p^{s_{2}}}^{*}$. Therefore,
\begin{equation*}
  \begin{split}
  N_{\alpha}&=\varepsilon p^{\frac{r}{2}-s_{1}-s_{2}}\sum_{z \in \mathbb{F}_{p^{s_{1}}}^{*}}\sum_{y \in \mathbb{F}_{p^{s_{2}}}^{*}}\zeta_{p}^{Tr_{1}^{m}(\lambda^{-e}yzF^{*}(\alpha))}+\varepsilon p^{\frac{r}{2}-s_{1}-s_{2}}(p^{s_{2}}-1)
   +p^{r-s_{1}-s_{2}}\\
  &=\varepsilon p^{\frac{r}{2}-s_{1}-s_{2}}\sum_{y \in \mathbb{F}_{p^{s_{2}}}^{*}}\sum_{z \in \mathbb{F}_{p^{s_{1}}}^{*}}\zeta_{p}^{Tr_{1}^{s_{1}}(z(Tr_{s_{1}}^{m}(\lambda^{-e}yF^{*}(\alpha)))}\\
  & \ \ +\varepsilon p^{\frac{r}{2}-s_{1}-s_{2}}(p^{s_{2}}-1)+p^{r-s_{1}-s_{2}}\\
  &=\varepsilon p^{\frac{r}{2}-s_{1}-s_{2}}\sum_{y \in \mathbb{F}_{p^{s_{2}}}^{*}}\sum_{z \in \mathbb{F}_{p^{s_{1}}}}\zeta_{p}^{Tr_{1}^{s_{1}}(z(Tr_{s_{1}}^{m}(\lambda^{-e}yF^{*}(\alpha)))}+p^{r-s_{1}-s_{2}}\\
  &=\varepsilon p^{\frac{r}{2}-s_{2}}|\{y \in \mathbb{F}_{p^{s_{2}}}^{*}: Tr_{s_{1}}^{m}(\lambda^{-e}yF^{*}(\alpha))=0\}|+p^{r-s_{1}-s_{2}}.\\
  \end{split}
\end{equation*}
When $Tr_{s_{2}}^{m}(\lambda^{-e}F^{*}(\alpha))=0$, then for any $y \in \mathbb{F}_{p^{s_{2}}}^{*}$ we have $Tr_{s_{1}}^{m}(\lambda^{-e}yF^{*}(\alpha))=Tr_{s_{1}}^{s_{2}}(y(Tr_{s_{2}}^{m}(\lambda^{-e}F^{*}(\alpha))))=0$. Thus in this case $|\{y \in \mathbb{F}_{p^{s_{2}}}^{*}: Tr_{s_{1}}^{m}(\lambda^{-e}yF^{*}(\alpha))$ $=0\}|=p^{s_{2}}-1$, and $N_{\alpha}=\varepsilon p^{\frac{r}{2}-s_{2}}(p^{s_{2}}-1)+p^{r-s_{1}-s_{2}}$. When $Tr_{s_{2}}^{m}(\lambda^{-e}F^{*}(\alpha))\neq0$, then  $Tr_{s_{1}}^{m}(\lambda^{-e}yF^{*}(\alpha))=Tr_{s_{1}}^{s_{2}}(y(Tr_{s_{2}}^{m}(\lambda^{-e}F^{*}(\alpha))))$ is a balanced function from $\mathbb{F}_{p^{s_{2}}}$ to $\mathbb{F}_{p^{s_{1}}}$. Thus in this case $|\{y \in \mathbb{F}_{p^{s_{2}}}^{*}: Tr_{s_{1}}^{m}(\lambda^{-e}yF^{*}(\alpha))=0\}|=p^{s_{2}-s_{1}}-1$, and $N_{\alpha}=\varepsilon p^{\frac{r}{2}-s_{2}}(p^{s_{2}-s_{1}}-1)+p^{r-s_{1}-s_{2}}$. By the above argument, for any $\alpha \neq (0, \dots, 0)$ we obtain
\begin{equation*}
  wt(c_{\alpha})=\begin{cases}
  p^{r-s_{1}-s_{2}}(p^{s_{1}}-1), & \text{ if } Tr_{s_{2}}^{m}(\lambda^{-e}F^{*}(\alpha))=0,\\
  (p^{r-s_{1}-s_{2}}+\varepsilon p^{\frac{r}{2}-s_{1}})(p^{s_{1}}-1), & \text{ if } Tr_{s_{2}}^{m}(\lambda^{-e}F^{*}(\alpha))\neq 0.
  \end{cases}
 \end{equation*}
It is easy to see that if $s_{2}\neq \frac{r}{2}$ when $\varepsilon=-1$, then $wt(c_{\alpha})=0$ if and only if $\alpha=(0, \dots, 0)$, that is, $c_{\alpha}=(0, \dots, 0)$ if and only if $\alpha=(0, \dots, 0)$. Hence, the number of codewords is $p^{r}$ and the dimension of linear code $\mathcal{C}_{D(F)}$ is $\frac{r}{s_{1}}$. In the following, we compute the values of $A_{p^{r-s_{1}-s_{2}}(p^{s_{1}}-1)}$ and $A_{(p^{r-s_{1}-s_{2}}+\varepsilon p^{\frac{r}{2}-s_{1}})(p^{s_{1}}-1)}$. With the similar argument as in the proof of Theorem 1, we have
$|\{\alpha \in V_{r}: Tr_{s_{2}}^{m}(\lambda^{-e}F^{*}(\alpha))=0\}|=p^{r-s_{2}}+\varepsilon p^{\frac{r}{2}-s_{2}}(p^{s_{2}}-1)$. Since $F^{*}(0)=0$, we have
\begin{equation*}
\begin{split}
  A_{p^{r-s_{1}-s_{2}}(p^{s_{1}}-1)}&=|\{\alpha \in V_{r}^{*}: Tr_{s_{2}}^{m}(\lambda^{-e}F^{*}(\alpha))=0\}|\\
  &=p^{r-s_{2}}+\varepsilon p^{\frac{r}{2}-s_{2}}(p^{s_{2}}-1)-1,\\
\end{split}
\end{equation*}
and
\begin{equation*}
\begin{split}
  A_{(p^{r-s_{1}-s_{2}}+\varepsilon p^{\frac{r}{2}-s_{1}})(p^{s_{1}}-1)}&=|\{\alpha \in V_{r}^{*}: Tr_{s_{2}}^{m}(\lambda^{-e}F^{*}(\alpha))\neq 0\}|\\
   &=p^{r}-1-A_{p^{r-s_{1}-s_{2}}(p^{s_{1}}-1)}\\
  &=(p^{r-s_{2}}-\varepsilon p^{\frac{r}{2}-s_{2}})(p^{s_{2}}-1). \ \ \ \Box\\
\end{split}
\end{equation*}

Keep the same notation as in Theorem 2. If $F$ with the Condition A also satisfies that $F(cx)=c^{l}F(x)$, $c\in \mathbb{F}_{p^m}^{*}, x \in V_{r}$ for some positive integer $l$, then it is easy to see that $Tr_{s_{2}}^{m}(\lambda F(x))=0$ if and only if $Tr_{s_{2}}^{m}(\lambda F(cx))=0$ for all $c \in \mathbb{F}_{p^{s_{1}}}^{*}$. Then we can select a subset $\widetilde{D}(F)=\{w^{[1]}, \dots, w^{[u]}\}$ of $D(F)$ such that
\begin{equation}\label{18}
  D(F)=\bigcup_{i=1}^{u}w^{[i]}\mathbb{F}_{p^{s_{1}}}^{*},
\end{equation}
where $u=\frac{p^{r-s_{2}}+\varepsilon p^{\frac{r}{2}-s_{2}}(p^{s_{2}}-1)-1}{p^{s_{1}}-1}$. Note that such a condition holds for the vectorial dual-bent functions listed in Section 2 (including bent functions belonging to $\mathcal{RF}$). The following corollary is directly from Theorem 2.

\begin{corollary}\label{Corollary1}
Let $F: V_{r} \rightarrow \mathbb{F}_{p^m}$ be a vectorial dual-bent function satisfying the Condition A and $F(cx)=c^{l}F(x), c\in \mathbb{F}_{p^m}^{*}, x \in V_{r}$ for some positive integer $l$. Let $s_{1}, s_{2}$ be positive integers with $s_{2} \mid m, s_{1} \mid s_{2}$, and $s_{2}\neq \frac{r}{2}$ if $\varepsilon=-1$. Let $q=p^{s_{1}}$, $\lambda \in \mathbb{F}_{p^m}^{*}$ and $\widetilde{D}(F)=\{w^{[1]}, \dots, w^{[u]}\}$ be a subset of $D(F)=\{x \in V_{r}^{*}: Tr_{s_{2}}^{m}(\lambda F(x))=0\}$ such that (12) holds. Define
\begin{equation}\label{13}
  \mathcal{C}_{\widetilde{D}(F)}=\{c_{\alpha}=(\sum_{j=1}^{t}Tr_{s_{1}}^{r_{j}}(\alpha_{j}d_{j}))_{d=(d_{1}, \dots, d_{t}) \in \widetilde{D}(F)}: \alpha=(\alpha_{1}, \dots, \alpha_{t})\in V_{r}\}.
\end{equation}
Then $\mathcal{C}_{\widetilde{D}(F)}$ defined by (13) is a two-weight $q$-ary linear code with parameters $[\frac{p^{r-s_{2}}+\varepsilon p^{\frac{r}{2}-s_{2}}(p^{s_{2}}-1)-1}{p^{s_{1}}-1}, \frac{r}{s_{1}}]$ and the weight distribution is given in Table 3.
\end{corollary}
\begin{table}\label{3}
  \caption{The weight distribution of $\mathcal{C}_{\widetilde{D}(F)}$ defined by (13)}
  \begin{tabular}{|c|c|}
    \hline
    Hamming weight $a$ & Multiplicity $A_{a}$ \\
    \hline
    \hline
    $0$ & $1$ \\
    $p^{r-s_{1}-s_{2}}$ & $p^{r-s_{2}}+\varepsilon p^{\frac{r}{2}-s_{2}}(p^{s_{2}}-1)-1$ \\
    $p^{r-s_{1}-s_{2}}+\varepsilon p^{\frac{r}{2}-s_{1}}$ & $(p^{s_{2}}-1)(p^{r-s_{2}}-\varepsilon p^{\frac{r}{2}-s_{2}})$ \\
    \hline
  \end{tabular}
\end{table}

\begin{remark}
As we illustrate before, any bent function $f \in \mathcal{RF}$ satisfies the conditions in Theorem 2 and Corollary 1. When $s_{1}=s_{2}=1$ (which corresponds to $p$-ary linear code), the linear codes constructed in Theorem 2 and Corollary 1 reduce to the linear codes constructed in Theorem 15 and Corollary 17 of \cite{Tang}.
\end{remark}

When $q\neq p$, based on vectorial dual-bent functions listed by (1)-(6), we list the corresponding two-weight $q$-ary linear codes from Theorem 2 and Corollary 1.
\begin{itemize}
  \item Let $F: \mathbb{F}_{p^{r'}} \times \mathbb{F}_{p^{r'}}\rightarrow \mathbb{F}_{p^m}$ be defined as (1) (respectively, (2), (5)). Then $F$ is vectorial dual-bent satisfying the conditions in Theorem 2 and Corollary 1, and the corresponding value $\varepsilon=1$. Let $s_{1}, s_{2}$ be positive integers with $s_{1} \mid s_{2}$ and $s_{2}\mid m$, and let $\lambda \in \mathbb{F}_{p^m}^{*}$. Define
      \begin{equation*}
       D(F)=\{x=(x_{1}, x_{2}) \in (\mathbb{F}_{p^{r'}}\times \mathbb{F}_{p^{r'}})^{*}: Tr_{s_{2}}^{r'}(\lambda a x_{1}x_{2}^{e})=0\},
      \end{equation*}
  (respectively,
  \begin{equation*}
  \begin{split}
    &D(F)=\{x=(x_{1}, x_{2}) \in (\mathbb{F}_{p^{r'}}\times \mathbb{F}_{p^{r'}})^{*}: Tr_{s_{2}}^{r'}(\lambda a x_{1}L(x_{2}))=0\},\\
    &D(F)=\{x=(x_{1}, x_{2})\in (\mathbb{F}_{p^{r'}} \times \mathbb{F}_{p^{r'}})^{*}: Tr_{s_{2}}^{m}(\lambda g(\gamma G(x_{1}x_{2}^{p^{r'}-2})))=0\}
  \end{split}
  \end{equation*}). Let $\widetilde{D}(F)=\{w^{[1]}, \dots, w^{[u]}\}$ such that (12) holds. Then the corresponding linear code $\mathcal{C}_{D(F)}$ defined by (11) is
  \begin{equation*}
  \begin{split}
  &\mathcal{C}_{D(F)}=\\
  &\{c_{\alpha}=(Tr_{s_{1}}^{r'}(\alpha_{1}d_{1}+\alpha_{2}d_{2}))_{d=(d_{1}, d_{2})\in D(F)}: \alpha=(\alpha_{1}, \alpha_{2}) \in \mathbb{F}_{p^{r'}}\times\mathbb{F}_{p^{r'}}\},
  \end{split}
  \end{equation*}
  which is a two-weight $p^{s_{1}}$-ary $[p^{2r'-s_{2}}+p^{r'-s_{2}}(p^{s_{2}}-1)-1, \frac{2r'}{s_{1}}]$ linear code.
  The corresponding linear code $\mathcal{C}_{\widetilde{D}(F)}$ defined by (13) is a two-weight $p^{s_{1}}$-ary $[\frac{p^{2r'-s_{2}}+ p^{r'-s_{2}}(p^{s_{2}}-1)-1}{p^{s_{1}}-1}, \frac{2r'}{s_{1}}]$ linear code.
  \item Let $F: \mathbb{F}_{p^{r}}\rightarrow \mathbb{F}_{p^m}$ be defined as (3) with $2m \mid r$. Then $F$ is vectorial dual-bent satisfying the conditions in Theorem 2 and Corollary 1, and the corresponding value $\varepsilon=-\epsilon^{{r}}\eta_{r}(a)$. Let $s_{1}, s_{2}$ be positive integers with $s_{1} \mid s_{2}$ and $s_{2}\mid m$ and $s_{2}\neq \frac{r}{2}$ if $\epsilon^{{r}}\eta_{r}(a)=1$, and let $\lambda \in \mathbb{F}_{p^m}^{*}$. Define $D(F)=\{x \in \mathbb{F}_{p^{r}}^{*}: Tr_{s_{2}}^{r}(\lambda ax^{2})=0\}$ and $\widetilde{D}(F)=\{w^{[1]}, \dots, w^{[u]}\}$ such that (12) holds. Then the corresponding linear codes $\mathcal{C}_{D(F)}$ defined by (11) and $\mathcal{C}_{\widetilde{D}(F)}$ defined by (13) are
  \begin{equation}\label{14}
  \begin{split}
  &\mathcal{C}_{D(F)}=\{c_{\alpha}=(Tr_{s_{1}}^{r}(\alpha d))_{d\in D(F)}: \alpha \in\mathbb{F}_{p^{r}}\}
  \end{split}
  \end{equation}
  and
  \begin{equation}\label{15}
  \begin{split}
  &\mathcal{C}_{\widetilde{D}(F)}=\{c_{\alpha}=(Tr_{s_{1}}^{r}(\alpha d))_{d\in \widetilde{D}(F)}: \alpha \in\mathbb{F}_{p^{r}}\},
  \end{split}
  \end{equation}
  respectively, which are two-weight $p^{s_{1}}$-ary $[p^{r-s_{2}}-\epsilon^{{r}}\eta_{r}(a) p^{\frac{r}{2}-s_{2}}(p^{s_{2}}-1)-1, \frac{r}{s_{1}}]$ linear code and two-weight $p^{s_{1}}$-ary $[\frac{p^{r-s_{2}}-\epsilon^{{r}}\eta_{r}(a) p^{\frac{r}{2}-s_{2}}(p^{s_{2}}-1)-1}{p^{s_{1}}-1}, \frac{r}{s_{1}}]$ linear code, respectively. In particular, when $s_{1}=1$ (which corresponds to $p$-ary linear code), the linear codes given by (14) and (15) reduce to the linear codes given in Theorem 2.2 and Corollary 2.4 of \cite{Tang2} respectively.
  \item Let $F: \mathbb{F}_{p^m}^{t}\rightarrow \mathbb{F}_{p^m}$ be defined as (4) with $t$ even. Then $F$ is vectorial dual-bent satisfying the conditions in Theorem 2 and Corollary 1, and the corresponding value $\varepsilon=\epsilon^{mt}\eta_{m}(a_{1}\cdots a_{t})$. Let $s_{1}, s_{2}$ be positive integers with $s_{1} \mid s_{2} \mid m$ and $s_{2}\neq \frac{mt}{2}$ if $\epsilon^{mt}\eta_{m}(a_{1}\cdots a_{t})=-1$, and let $\lambda \in \mathbb{F}_{p^m}^{*}$. Define $D(F)=\{x=(x_{1}, \dots, x_{t}) \in (\mathbb{F}_{p^m}^{t})^{*}: \sum_{i=1}^{t}Tr_{s_{2}}^{m}(\lambda a_{i}x_{i}^{2})=0\}$ and $\widetilde{D}(F)=\{w^{[1]}, \dots, w^{[u]}\}$ such that (13) holds. Then the corresponding linear code $\mathcal{C}_{D(F)}$ defined by (11) is
  \begin{equation*}
  \begin{split}
  &\mathcal{C}_{D(F)}=\{c_{\alpha}=(\sum_{i=1}^{t}Tr_{s_{1}}^{m}(\alpha_{j}d_{j}))_{d=(d_{1}, \dots, d_{t})\in D(F)}: \alpha=(\alpha_{1}, \dots, \alpha_{t}) \in\mathbb{F}_{p^m}^{t}\},
  \end{split}
  \end{equation*}
  which is a two-weight $p^{s_{1}}$-ary $[p^{mt-s_{2}}+\epsilon^{mt}\eta_{m}(a_{1}\cdots a_{t})p^{\frac{mt}{2}-s_{2}}(p^{s_{2}}-1)-1, \frac{mt}{s_{1}}]$ linear code.
  The corresponding linear code $C_{\widetilde{D}(F)}$ defined by (13) is a two-weight $p^{s_{1}}$-ary $[\frac{p^{mt-s_{2}}+\epsilon^{mt}\eta_{m}(a_{1}\cdots a_{t})p^{\frac{mt}{2}-s_{2}}(p^{s_{2}}-1)-1}{p^{s_{1}}-1}, \frac{mt}{s_{1}}]$ linear code.
  \item  Let $F: \mathbb{F}_{p^{r'}} \times \mathbb{F}_{p^{r''}} \times \mathbb{F}_{p^{r''}}\rightarrow \mathbb{F}_{p^m}$ be defined as (6) for which $2m \mid r'$ and $\alpha_{j}, 1\leq j \leq 3$ are all squares or non-squares in $\mathbb{F}_{p^{r'}}^{*}$. Then $F$ is vectorial dual-bent satisfying the conditions in Theorem 2 and Corollary 1, and the corresponding value $\varepsilon=-\epsilon^{r'}\eta_{r'}(\alpha_{1})$. Let $s_{1}, s_{2}$ be positive integers with $s_{1} \mid s_{2}$ and $s_{2}\mid m$, and let $\lambda \in \mathbb{F}_{p^m}^{*}$. Define $D(F)=\{x=(x_{1}, x_{2}, x_{3}) \in (\mathbb{F}_{p^{r'}}\times \mathbb{F}_{p^{r''}}\times \mathbb{F}_{p^{r''}})^{*}: Tr_{s_{2}}^{m}(\lambda F(x_{1}, x_{2}, x_{3}))=0\}$ and $\widetilde{D}(F)=\{w^{[1]}, \dots, w^{[u]}\}$ such that (13) holds. Then the corresponding linear code $\mathcal{C}_{D(F)}$ defined by (11) is
  \begin{equation*}
  \begin{split}
  &\mathcal{C}_{D(F)}=\{c_{\alpha}=(Tr_{s_{1}}^{r'}(\alpha_{1}d_{1})+Tr_{s_{1}}^{r''}(\alpha_{2}d_{2}+\alpha_{3}d_{3}))_{d=(d_{1}, d_{2}, d_{3}) \in D(F)}: \\
  &\ \ \ \ \ \ \ \ \ \ \ \ \ \alpha=(\alpha_{1}, \alpha_{2}, \alpha_{3}) \in \mathbb{F}_{p^{r'}}\times \mathbb{F}_{p^{r''}}\times \mathbb{F}_{p^{r''}}\},
  \end{split}
  \end{equation*}
  which is a two-weight $p^{s_{1}}$-ary $[p^{r'+2r''-s_{2}}-\epsilon^{r'}\eta_{r'}(\alpha_{1})p^{\frac{r'}{2}+r''-s_{2}}(p^{s_{2}}-1)-1, \frac{r'+2r''}{s_{1}}]$ linear code.
  The corresponding linear code $C_{\widetilde{D}(F)}$ defined by (13) is a two-weight $p^{s_{1}}$-ary $[\frac{p^{r'+2r''-s_{2}}-\epsilon^{r'}\eta_{r'}(\alpha_{1}) p^{\frac{r'}{2}+r''-s_{2}}(p^{s_{2}}-1)-1}{p^{s_{1}}-1}, \frac{r'+2r''}{s_{1}}]$ linear code.
\end{itemize}

Keep the same notation as in Corollary 1. We illustrate that the constructed linear code from Corollary 1 can be used to construct optimal codes in a special case. When $s_{1}=s_{2}=\frac{r}{4}$ and $\varepsilon=-1$, then it is easy to verify that the $p^{\frac{r}{4}}$-ary linear code $\mathcal{C}_{\widetilde{D}(F)}$ constructed in Corollary 1 has the parameters $[p^{\frac{r}{2}}+1, 4, p^{\frac{r}{4}}(p^{\frac{r}{4}}-1)]$ which meets the Griesmer bound. We list the vectorial dual-bent functions $F: V_{r}\rightarrow \mathbb{F}_{p^m}$ which make the corresponding linear codes $\mathcal{C}_{\widetilde{D}(F)}$ defined by (13) meet the Griesmer bound when $s_{1}=s_{2}=\frac{r}{4}$:

\begin{itemize}
  \item Let $r\geq 4$ be an integer with $4 \mid r$, $m=\frac{r}{4}$ or $\frac{r}{2}$, and $a \in \mathbb{F}_{p^r}^{*}$ be a square. Define
  $F: \mathbb{F}_{p^r}\rightarrow \mathbb{F}_{p^m}$ as
  \begin{equation*}
  F(x)=Tr_{m}^{r}(ax^{2}).
  \end{equation*}
  \item Let $t, m, r$ be positive integers for which $4 \mid r$, $m=\frac{r}{4}$ or $\frac{r}{2}$, $t=\frac{r}{m}$, and $a_{1}, \dots, a_{t}\in \mathbb{F}_{p^m}^{*}$ for which $\prod_{i=1}^{t}a_{i}$ is a non-square in $\mathbb{F}_{p^m}^{*}$. Define $F: \mathbb{F}_{p^m}^{t}\rightarrow \mathbb{F}_{p^m}$ as
  \begin{equation*}
    F(x)=a_{1}x_{1}^{2}+\dots+a_{t}x_{t}^{2}, x=(x_{1}, \dots, x_{t})\in \mathbb{F}_{p^m}^{t}.
  \end{equation*}
  \item Let $m, r$ be positive integers with $4 \mid r, m=\frac{r}{4}$, and $\beta \in \mathbb{F}_{p^m}^{*}, a_{1}, a_{2}\in \mathbb{F}_{p^{2m}}^{*}$ with $\eta_{2m}(a_{i})=\epsilon^{2m}$ for $i=1, 2$. Define $F: \mathbb{F}_{p^{2m}}\times \mathbb{F}_{p^m}\times \mathbb{F}_{p^m}\rightarrow \mathbb{F}_{p^m}$ as
      \begin{equation*}
      \begin{split}
        &F(x)=x_{3}^{p^m-1}Tr_{m}^{2m}((a_{2}-a_{1})x_{1}^{2})+Tr_{m}^{2m}(a_{1}x_{1}^{2})+\beta x_{2}x_{3}, \\
        & x=(x_{1}, x_{2}, x_{3})\in \mathbb{F}_{p^{2m}} \times \mathbb{F}_{p^m} \times \mathbb{F}_{p^m}.\\
      \end{split}
      \end{equation*}
\end{itemize}

We give some examples.

\begin{example}\label{Example2}
Let $p=3, s_{1}=s_{2}=2, m=4, r=8$ and $\lambda=1$, $F: \mathbb{F}_{3^8}\rightarrow \mathbb{F}_{3^4}$ be defined as $F(x)=Tr_{4}^{8}(x^{2})$. The linear code $\mathcal{C}_{\widetilde{D}(F)}$ defined by (13) is
\begin{equation*}
  \mathcal{C}_{\widetilde{D}(F)}=\{c_{\alpha}=(Tr_{2}^{8}(\alpha d))_{d \in \widetilde{D}(F)}: \alpha \in \mathbb{F}_{3^8}\},
\end{equation*}
where $\widetilde{D}(F)=\{w^{[1]}, \dots, w^{[82]}\}$ such that $D(F)=\bigcup_{i=1}^{82}w^{[i]}\mathbb{F}_{3^2}^{*}$, where $D(F)=\{x \in \mathbb{F}_{3^8}^{*}: Tr_{2}^{8}(x^2)=0\}$. Then $\mathcal{C}_{\widetilde{D}(F)}$ is a two-weight $9$-ary linear code with parameters $[82, 4, 72]$ which meets the Griesmer bound and the weight enumerator is $1+5904z^{72}+656z^{81}$.
\end{example}

\begin{example}\label{Example3}
Let $p=5, s_{1}=s_{2}=m=2, r=8$ and $\lambda=1$, $F: \mathbb{F}_{5^2}^{4}\rightarrow \mathbb{F}_{5^2}$ be defined as $F(x)=x_{1}^{2}+x_{2}^{2}+x_{3}^{2}+\gamma x_{4}^{2}$, where $\gamma$ is a primitive element of $\mathbb{F}_{5^2}$. The linear code  $\mathcal{C}_{\widetilde{D}(F)}$ defined by (13) is
\begin{equation*}
  \mathcal{C}_{\widetilde{D}(F)}=\{c_{\alpha}=(\sum_{i=1}^{4}\alpha_{i}d_{i})_{d=(d_{1}, \dots, d_{4}) \in \widetilde{D}(F)}: \alpha=(\alpha_{1}, \dots, \alpha_{4}) \in \mathbb{F}_{5^2}^{4}\},
\end{equation*}
where $\widetilde{D}(F)=\{w^{[1]}, \dots, w^{[626]}\}$ such that $D(F)=\bigcup_{i=1}^{626}w^{[i]}\mathbb{F}_{5^2}^{*}$, where $D(F)=\{x=(x_{1}, \dots, x_{4}) $ $\in (\mathbb{F}_{5^2}^{4})^{*}: x_{1}^{2}+x_{2}^{2}+x_{3}^{2}+\gamma x_{4}^{2}=0\}$. Then $\mathcal{C}_{\widetilde{D}(F)}$ is a two-weight $25$-ary linear code with parameters $[626, 4, 600]$ which meets the Griesmer bound and the weight enumerator is $1+375600z^{600}+15024z^{625}$.
\end{example}

\begin{example}\label{Example4}
Let $p=7, s_{1}=s_{2}=m=2, r=8$ and $\lambda=1$, $F: \mathbb{F}_{7^4} \times \mathbb{F}_{7^2} \times \mathbb{F}_{7^2} \rightarrow \mathbb{F}_{7^2}$ be defined as $F(x)=x_{3}^{48}Tr_{2}^{4}((\gamma^2-1)x_{1}^{2})+Tr_{2}^{4}(x_{1}^{2})+x_{2}x_{3}, x=(x_{1}, x_{2}, x_{3}) \in \mathbb{F}_{7^4} \times \mathbb{F}_{7^2} \times \mathbb{F}_{7^2}$, where $\gamma$ is a primitive element of $\mathbb{F}_{7^4}$. The linear code $\mathcal{C}_{\widetilde{D}(F)}$ defined by (13) is
\begin{equation*}
\begin{split}
  & \mathcal{C}_{\widetilde{D}(F)}=\{c_{\alpha}=(Tr_{2}^{4}(\alpha_{1}d_{1})+\alpha_{2}d_{2}+\alpha_{3}d_{3})_{d=(d_{1}, d_{2}, d_{3}) \in \widetilde{D}(F)}: \\
  & \ \ \ \ \ \ \ \ \ \ \ \ \alpha=(\alpha_{1}, \alpha_{2}, \alpha_{3}) \in \mathbb{F}_{7^4} \times \mathbb{F}_{7^2} \times \mathbb{F}_{7^2}\},\\
  \end{split}
\end{equation*}
where $\widetilde{D}(F)=\{w^{[1]}, \dots, w^{[2402]}\}$ such that $D(F)=\bigcup_{i=1}^{2402}w^{[i]}\mathbb{F}_{7^2}^{*}$, where $D(F)=\{x=(x_{1}, x_{2}, $ $x_{3})\in (\mathbb{F}_{7^4} \times \mathbb{F}_{7^2} \times \mathbb{F}_{7^2})^{*}: x_{3}^{48}Tr_{2}^{4}((\gamma^2-1)x_{1}^{2})+Tr_{2}^{4}(x_{1}^{2})+x_{2}x_{3}=0\}$. Then $\mathcal{C}_{\widetilde{D}(F)}$ is a two-weight $49$-ary linear code with parameters $[2402, 4, 2352]$ which meets the Griesmer bound and the weight enumerator is $1+5649504z^{2352}$ $+115296z^{2401}$.
\end{example}
Based on Lemmas 2 and 4, we have the following proposition. For the convenience of expression, we both use $\mathcal{C}$ to represent the linear codes constructed in Theorem 2 and Corollary 1.

\begin{proposition}\label{Proposition4}
Let $\mathcal{C}$ be the $p^{s_{1}}$-ary linear code constructed in Theorem 2 or Corollary 1 with $s_{1}+s_{2}\leq \frac{r}{2}$ and $s_{1}+s_{2}\neq \frac{r}{2}$ if $\varepsilon=-1$, and let $G=[g_{0}, g_{1}, \dots, g_{n-1}]$ be its generator matrix. In the secret sharing scheme based on $\mathcal{C}^{\bot}$, there are altogether $p^{r-s_{1}}$ minimal access sets.
\begin{itemize}
  \item If $\mathcal{C}$ is constructed in Theorem 2:
     For any $1\leq i \leq n-1$, if $g_{i}$ is a multiple of $g_{0}$, then the participant $P_{i}$ must be in every minimal access set. If $g_{i}$ is not a multiple of $g_{0}$, then the participant $P_{i}$ must be in $(p^{s_{1}}-1)p^{r-2s_{1}}$ out of $p^{r-s_{1}}$ minimal access sets.
  \item If $\mathcal{C}$ is constructed in Corollary 1:
     For any fixed $1\leq l \leq min\{\frac{r}{s_{1}}-1, d^{\bot}-2\}$, every set of $l$ participants is involved in $p^{r-s_{1}(l+1)}(p^{s_{1}}-1)^{l}$ out of $p^{r-s_{1}}$ minimal access sets, where $d^{\bot}$ denotes the minimum distance of $\mathcal{C}^{\bot}$.
\end{itemize}
\end{proposition}
\textbf{Proof.}
From Lemma 2, it is easy to verify that the constructed $p^{s_{1}}$-ary linear code of Theorem 2 or Corollary 1 is minimal when $s_{1}+s_{2}\leq \frac{r}{2}$ and $s_{1}+s_{2}\neq \frac{r}{2}$ if $\varepsilon=-1$. If $\mathcal{C}$ is constructed in Theorem 2, then obviously $d^{\bot} \neq 1$ and from $F(x)=F(-x)$ we have $-D(F)=D(F)$, which implies that $d^{\bot}=2$. If $\mathcal{C}$ is constructed in Corollary 1, then obviously $d^{\bot} \neq 1$ and if $d^{\bot}=2$, then there exist $d=(d_{1}, \dots, d_{t}), d'=(d'_{1}, \dots, d'_{t})\in \widetilde{D}(F)$ and $z \in \mathbb{F}_{p^{s_{1}}}^{*}$ such that $\sum_{j=1}^{t}Tr_{s_{1}}^{r_{j}}(\alpha_{j}d_{j})+z\sum_{j=1}^{t}Tr_{s_{1}}^{r_{j}}(\alpha_{j}d'_{j})=0$ for all $\alpha=(\alpha_{1}, \dots, \alpha_{t})\in V_{r}$, which implies that $d+zd'=0$ and this is impossible by the definition of $\widetilde{D}(F)$ given by (13). Therefore, $d^{\bot}\geq 3$ if $\mathcal{C}$ is constructed in Corollary 1. Then the conclusion follows from Lemma 4. \ $\Box$

\begin{theorem}\label{Theorem3}
Let $F: V_{r} \rightarrow \mathbb{F}_{p^m}$ be a vectorial dual-bent function satisfying the Condition A. Let $s_{1}, s_{2}$ be positive integers with $s_{2} \mid m, s_{1} \mid s_{2}$. Let $q=p^{s_{1}}$, $\lambda \in \mathbb{F}_{p^m}^{*}$ and $S=\{x^{2}: x \in \mathbb{F}_{p^{s_{2}}}^{*}\}$, $N=\mathbb{F}_{p^{s_{2}}}^{*} \backslash S$. Define
\begin{equation}\label{16}
  \mathcal{C}_{D(F)}=\{c_{\alpha}=(\sum_{j=1}^{t}Tr_{s_{1}}^{r_{j}}(\alpha_{j}d_{j}))_{d=(d_{1}, \dots, d_{t}) \in D(F)}: \alpha=(\alpha_{1}, \dots, \alpha_{t})\in V_{r}\},
\end{equation}
where $D(F)=\{x \in V_{r}^{*}: Tr_{s_{2}}^{m}(\lambda F(x)) \in S\}$ (or $D(F)=\{x \in V_{r}^{*}: Tr_{s_{2}}^{m}(\lambda F(x)) \in N\}$). Then $\mathcal{C}_{D(F)}$ defined by (16) is a two-weight $q$-ary linear code with parameters $[\frac{(p^{s_{2}}-1)}{2}(p^{r-s_{2}}-\varepsilon p^{\frac{r}{2}-s_{2}}), \frac{r}{s_{1}}]$ and the weight distribution is given in Table 4.
\end{theorem}
\begin{table}\label{4}
  \caption{The weight distribution of $\mathcal{C}_{D(F)}$ defined by (16)}
  \begin{tabular}{|c|c|}
    \hline
    Hamming weight $a$ & Multiplicity $A_{a}$ \\
    \hline
    \hline
    $0$ & $1$ \\
    $\frac{ (p^{s_{2}}-1) (p^{s_{1}}-1)p^{r-s_{1}-s_{2}}
    	\ +\varepsilon (\epsilon^{2s_2}-1) p^{\frac{r}{2}-s_1} (p^{s_1}-1)}{2}$ & $\frac{p^{s_{2}}-1}{2}(p^{r-s_{2}}+\varepsilon p^{\frac{r}{2}-s_{2}})+p^{r-s_{2}}-1$ \\
    $\frac{(p^{s_{2}}-1)(p^{s_{1}}-1)p^{r-s_{1}-s_{2}}
    	\ -\varepsilon(\epsilon^{2s_2}+1)p^{\frac{r}{2}-s_1}(p^{s_1}-1)}{2}$ & $\frac{p^{s_{2}}-1}{2}(p^{r-s_{2}}-\varepsilon p^{\frac{r}{2}-s_{2}})$ \\
    \hline
  \end{tabular}
\end{table}
\textbf{Proof.}
For the simplicity, we only prove the case of $D(F)=\{x \in V_{r}^{*}: Tr_{s_{2}}^{m}(\lambda F(x))$ $\in S\}$ since the proof of the other case is similar. Note that $Tr_{s_{2}}^{m}(\lambda F(x))\in S$ if and only if $F(x)=\lambda^{-1}z_{1}^{[i]}, \dots, \lambda^{-1}z_{p^{m-s_{2}}}^{[i]}, 1\leq i\leq \frac{p^{s_{2}}-1}{2}$, where $z_{1}^{[i]}, \dots, z_{p^{m-s_{2}}}^{[i]}$ are the solutions of $Tr_{s_{2}}^{m}(z)=w^{[i]}$ and $S=\{w^{[1]}, \dots, w^{[\frac{p^{s_{2}}-1}{2}]}\}$. By Lemma 1, for any $a \in \mathbb{F}_{p^m}^{*}$ we have $|\{x \in V_{r}: F(x)=a\}|=p^{r-m}-\varepsilon p^{\frac{r}{2}-m}$. Since $F(0)=0$, we have $|\{x \in V_{r}^{*}: Tr_{s_{2}}^{m}(\lambda F(x))\in S\}|=
\frac{p^{s_{2}}-1}{2}p^{m-s_{2}}(p^{r-m}-\varepsilon p^{\frac{r}{2}-m})=\frac{p^{s_{2}}-1}{2}(p^{r-s_{2}}-\varepsilon p^{\frac{r}{2}-s_{2}})$, that is, the length of linear code $\mathcal{C}_{D(F)}$ is $|D(F)|=\frac{p^{s_{2}}-1}{2}(p^{r-s_{2}}-\varepsilon p^{\frac{r}{2}-s_{2}})$. When $\alpha=(0, \dots, 0)$, then $wt(c_{\alpha})=0$. When $\alpha=(\alpha_{1}, \dots, \alpha_{t})\neq (0, \dots, 0)$, then $wt(c_{\alpha})=|\{x \in V_{r}: Tr_{s_{2}}^{m}(\lambda F(x))\in S\}|-N_{\alpha}$, where
\begin{equation*}
\begin{split}
 N_{\alpha}&=|\{x=(x_{1}, \dots, x_{t}) \in V_{r}: Tr_{s_{2}}^{m}(\lambda F(x))\in S, \sum_{j=1}^{t}Tr_{s_{1}}^{r_{j}}(\alpha_{j}x_{j})=0\}|\\
&=p^{-s_{1}-s_{2}} \sum_{x=(x_{1}, \dots, x_{t}) \in V_{r}}\sum_{j \in S}\sum_{y \in \mathbb{F}_{p^{s_{2}}}}\zeta_{p}^{Tr_{1}^{s_{2}}(y(Tr_{s_{2}}^{m}(\lambda F(x))-j))}\\
& \ \ \ \times \sum_{z \in \mathbb{F}_{p^{s_{1}}}}\zeta_{p}^{Tr_{1}^{s_{1}}(-z(\sum_{j=1}^{t}Tr_{s_{1}}^{r_{j}}(\alpha_{j}x_{j})))}\\
&=p^{-s_{1}-s_{2}} \sum_{x=(x_{1}, \dots, x_{t}) \in V_{r}}\sum_{j \in S}\sum_{y \in \mathbb{F}_{p^{s_{2}}}^{*}}\zeta_{p}^{Tr_{1}^{m}(y\lambda F(x))-Tr_{1}^{s_{2}}(jy)}\\
 & \ \ \times \sum_{z \in \mathbb{F}_{p^{s_{1}}}}\zeta_{p}^{-\sum_{j=1}^{t}Tr_{1}^{r_{j}}(z\alpha_{j}x_{j})}+p^{-s_{1}-s_{2}}\cdot \frac{p^{s_{2}}-1}{2}\cdot p^{s_{1}}\cdot p^{r-s_{1}}\\
 &=p^{-s_{1}-s_{2}}\sum_{z \in \mathbb{F}_{p^{s_{1}}}}\sum_{y \in \mathbb{F}_{p^{s_{2}}}^{*}}\sum_{j \in S}\zeta_{p}^{-Tr_{1}^{s_{2}}(jy)}W_{F_{y\lambda}}(z\alpha)+\frac{p^{r-s_{1}-s_{2}}(p^{s_{2}}-1)}{2}\\
 &=\varepsilon p^{\frac{r}{2}-s_{1}-s_{2}}\sum_{z \in \mathbb{F}_{p^{s_{1}}}^{*}}\sum_{y \in \mathbb{F}_{p^{s_{2}}}^{*}}\sum_{j \in S}\zeta_{p}^{-Tr_{1}^{s_{2}}(jy)+Tr_{1}^{m}(\sigma(y\lambda)F^{*}(z\alpha))}\\
\end{split}
\end{equation*}
\begin{equation*}
\begin{split}
 & \ \ +\varepsilon p^{\frac{r}{2}-s_{1}-s_{2}}\frac{p^{s_{2}}-1}{2}\cdot (-1)+\frac{p^{r-s_{1}-s_{2}}(p^{s_{2}}-1)}{2}\\
 &=\varepsilon p^{\frac{r}{2}-s_{1}-s_{2}}\sum_{z \in \mathbb{F}_{p^{s_{1}}}^{*}}\sum_{j \in S}\sum_{y \in \mathbb{F}_{p^{s_{2}}}^{*}}\zeta_{p}^{-Tr_{1}^{s_{2}}(jy)+Tr_{1}^{m}(\lambda^{-e}(\frac{z}{y})^{e}zF^{*}(\alpha))}\\
 & \ \ +\frac{(p^{s_{2}}-1)(p^{r-s_{1}-s_{2}}-\varepsilon p^{\frac{r}{2}-s_{1}-s_{2}})}{2},\\
\end{split}
\end{equation*}
where the last equation is obtained by $\sigma(c)=c^{-e}, F^{*}(cx)=c^{e+1}F^{*}(x), c \in \mathbb{F}_{p^m}^{*}, x \in V_{r}$ with $gcd(e, p^m-1)=1$.
For any fixed $z \in \mathbb{F}_{p^{s_{1}}}^{*}$, since $s_{1} \mid s_{2}$, we have that $P(y)=zy$ is a permutation over $\mathbb{F}_{p^{s_{2}}}^{*}$. Therefore,
\begin{equation*}
\begin{split}
  N_{\alpha}&=\varepsilon p^{\frac{r}{2}-s_{1}-s_{2}}\sum_{z \in \mathbb{F}_{p^{s_{1}}}^{*}}\sum_{j \in S}\sum_{y \in \mathbb{F}_{p^{s_{2}}}^{*}}\zeta_{p}^{-Tr_{1}^{s_{2}}(jyz)+Tr_{1}^{m}(\lambda^{-e}y^{-e}zF^{*}(\alpha))}\\
 & \ \ +\frac{(p^{s_{2}}-1)(p^{r-s_{1}-s_{2}}-\varepsilon p^{\frac{r}{2}-s_{1}-s_{2}})}{2}\\
 &=\varepsilon p^{\frac{r}{2}-s_{1}-s_{2}}\sum_{y \in \mathbb{F}_{p^{s_{2}}}^{*}}\sum_{z \in \mathbb{F}_{p^{s_{1}}}^{*}}\zeta_{p}^{Tr_{1}^{m}(-\lambda^{-e}y^{-e}zF^{*}(\alpha))}\sum_{j \in S}\zeta_{p}^{Tr_{1}^{s_{2}}(jyz)}\\
& \ \ +\frac{(p^{s_{2}}-1)(p^{r-s_{1}-s_{2}}-\varepsilon p^{\frac{r}{2}-s_{1}-s_{2}})}{2}.\\
\end{split}
\end{equation*}
By Lemma 3, we have
\begin{equation*}
  \begin{split}
   &\sum_{y \in \mathbb{F}_{p^{s_{2}}}^{*}}\sum_{z \in \mathbb{F}_{p^{s_{1}}}^{*}}\zeta_{p}^{Tr_{1}^{m}(-\lambda^{-e}y^{-e}zF^{*}(\alpha))}\sum_{j \in S}\zeta_{p}^{Tr_{1}^{s_{2}}(jyz)}\\
   &=\sum_{z \in \mathbb{F}_{p^{s_{1}}}^{*}}\sum_{\substack{y \in \mathbb{F}_{p^{s_{2}}}^{*}\\zy \in S}}\zeta_{p}^{Tr_{1}^{m}(-\lambda^{-e}y^{-e}zF^{*}(\alpha))}C_{0}
    +\sum_{z \in \mathbb{F}_{p^{s_{1}}}^{*}}\sum_{\substack{y \in \mathbb{F}_{p^{s_{2}}}^{*}\\zy \in N}}\zeta_{p}^{Tr_{1}^{m}(-\lambda^{-e}y^{-e}zF^{*}(\alpha))}C_{1},\\
\end{split}
\end{equation*}
where $C_{0}=\frac{(-1)^{s_{2}-1}\epsilon^{s_{2}}p^{\frac{s_{2}}{2}}-1}{2}$, $C_{1}=\frac{-(-1)^{s_{2}-1}\epsilon^{s_{2}}p^{\frac{s_{2}}{2}}-1}{2}$. Since $s_{1}\mid s_{2}$ and $s_{2}\mid m$ and $gcd(e, p^m-1)=1$, we can see that for any fixed $z \in \mathbb{F}_{p^{s_{1}}}^{*}$, $zy \in S$ if and only if $zy^{-e} \in S$. Therefore,
\begin{equation*}
\begin{split}
&\sum_{y \in \mathbb{F}_{p^{s_{2}}}^{*}}\sum_{z \in \mathbb{F}_{p^{s_{1}}}^{*}}\zeta_{p}^{Tr_{1}^{m}(-\lambda^{-e}y^{-e}zF^{*}(\alpha))}\sum_{j \in S}\zeta_{p}^{Tr_{1}^{s_{2}}(jyz)}\\
\end{split}
\end{equation*}
\begin{equation*}
\begin{split}
&=\sum_{z \in \mathbb{F}_{p^{s_{1}}}^{*}}\sum_{y \in S}\zeta_{p}^{Tr_{1}^{m}(-\lambda^{-e}yF^{*}(\alpha))}C_{0}+\sum_{z \in \mathbb{F}_{p^{s_{1}}}^{*}}\sum_{y\in N}\zeta_{p}^{Tr_{1}^{m}(-\lambda^{-e}yF^{*}(\alpha))}C_{1}\\
&=(p^{s_{1}}-1)((C_{0}-C_{1})\sum_{y \in S}\zeta_{p}^{Tr_{1}^{s_{2}}(yTr_{s_{2}}^{m}(-\lambda^{-e}F^{*}(\alpha)))}\\
& \ \ \ \ \ \ \ \ \ \ \ \ \ \ \ \ +C_{1}\sum_{y \in \mathbb{F}_{p^{s_{2}}}^{*}}\zeta_{p}^{Tr_{1}^{s_{2}}(yTr_{s_{2}}^{m}(-\lambda^{-e}F^{*}(\alpha)))})\\
&=\begin{cases}
\frac{(p^{s_{1}}-1)(1-\epsilon^{2s_{2}}p^{s_{2}})}{2}, & \text{ if } Tr_{s_{2}}^{m}(-\lambda^{-e}F^{*}(\alpha))\notin S,\\
\frac{(p^{s_{1}}-1)(\epsilon^{2s_{2}}p^{s_{2}}+1)}{2}, & \text{ if } Tr_{s_{2}}^{m}(-\lambda^{-e}F^{*}(\alpha))\in S.\\
\end{cases}
\end{split}
\end{equation*}
Then for any $\alpha\neq (0, \dots, 0)$, we have
\begin{equation*}
\begin{split}
  &N_{\alpha}=\\
  &\begin{cases}
  \frac{p^{r-s_1-s_2}(p^{s_2}-1)+\varepsilon(1-\epsilon^{2s_{2}}p^{s_2})p^{\frac{r}{2}-s_2}+\varepsilon p^{\frac{r}{2}-s_1}(\epsilon^{2s_{2}}-1) }{2},  \ \ \text {if } Tr_{s_{2}}^{m}(-\lambda^{-e}F^{*}(\alpha))\notin S,\\
  \frac{p^{r-s_1-s_2}(p^{s_2}-1)+\varepsilon(1+\epsilon^{2s_{2}}p^{s_2})p^{\frac{r}{2}-s_2}-\varepsilon p^{\frac{r}{2}-s_1}(\epsilon^{2s_{2}}+1) }{2},   \ \   \text {if } Tr_{s_{2}}^{m}(-\lambda^{-e}F^{*}(\alpha))\in S,\\
  \end{cases}
  \end{split}
\end{equation*}
and
\begin{equation}\label{17}
\begin{split}
 & wt(c_{\alpha})=\\
 &\begin{cases}
  \frac{ (p^{s_{2}}-1) (p^{s_{1}}-1)p^{r-s_{1}-s_{2}}
  	\ +\varepsilon (\epsilon^{2s_2}-1) p^{\frac{r}{2}-s_1} (p^{s_1}-1)}{2}, \ \text{ if } Tr_{s_{2}}^{m}(-\lambda^{-e}F^{*}(\alpha))\notin S,\\
  \frac{(p^{s_{2}}-1)(p^{s_{1}}-1)p^{r-s_{1}-s_{2}}
   \ -\varepsilon(\epsilon^{2s_2}+1)p^{\frac{r}{2}-s_1}(p^{s_1}-1)}{2}, \ \text{ if } Tr_{s_{2}}^{m}(-\lambda^{-e}F^{*}(\alpha))\in S.\\
  \end{cases}
  \end{split}
\end{equation}
By (17), we can see that $wt(c_{\alpha})=0$ if and only if $\alpha=(0, \dots, 0)$, that is, $c_{\alpha}=(0, \dots, 0)$ if and only if $\alpha=(0, \dots, 0)$. Thus, the dimension of $\mathcal{C}_{D(F)}$ is $\frac{r}{s_{1}}$. With the similar argument as in the proof of Theorem 1, we have $|\{x \in V_{r}: Tr_{s_{2}}^{m}(-\lambda^{-e}F^{*}(x))\in S\}|=\frac{p^{s_{2}}-1}{2}\cdot p^{m-s_{2}}\cdot (p^{r-m}-\varepsilon p^{\frac{r}{2}-m})=\frac{p^{s_{2}}-1}{2}(p^{r-s_{2}}-\varepsilon p^{\frac{r}{2}-s_{2}})$. Since $F^{*}(0)=0$, then
\begin{equation*}
\begin{split}
  &A_{\frac{(p^{s_{2}}-1)(p^{s_{1}}-1)p^{r-s_{1}-s_{2}}
  		\ -\varepsilon(\epsilon^{2s_2}+1)p^{\frac{r}{2}-s_1}(p^{s_1}-1)}{2}}\\
  &=\frac{p^{s_{2}}-1}{2}(p^{r-s_{2}}-\varepsilon p^{\frac{r}{2}-s_{2}}),\\
  \end{split}
\end{equation*}
and
\begin{equation*}
\begin{split}
 & A_{\frac{ (p^{s_{2}}-1) (p^{s_{1}}-1)p^{r-s_{1}-s_{2}}
  		\ +\varepsilon (\epsilon^{2s_2}-1) p^{\frac{r}{2}-s_1} (p^{s_1}-1)}{2}}\\
  &=\frac{p^{s_{2}}-1}{2}(p^{r-s_{2}}+\varepsilon p^{\frac{r}{2}-s_{2}})+p^{r-s_{2}}-1. \ \ \ \Box\\
  \end{split}
\end{equation*}

Keep the same notation as in Theorem 3. If $F$ with the Condition A also satisfies that $F(cx)=c^{l}F(x)$, $c\in \mathbb{F}_{p^m}^{*}, x \in V_{r}$ for some even positive integer $l$, then it is easy to see that $Tr_{s_{2}}^{m}(\lambda F(x))\in S$ (resp., $Tr_{s_{2}}^{m}(\lambda F(x))\in N$) if and only if $Tr_{s_{2}}^{m}(\lambda F(cx))\in S$ (resp., $Tr_{s_{2}}^{m}(\lambda F(cx))\in N$) for all $c \in \mathbb{F}_{p^{s_{1}}}^{*}$. Then we can select a subset $\widetilde{D}(F)=\{w^{[1]}, \dots, w^{[u]}\}$ of $D(F)$ such that
\begin{equation}\label{18}
  D(F)=\bigcup_{i=1}^{u}w^{[i]}\mathbb{F}_{p^{s_{1}}}^{*},
\end{equation}
where $u=\frac{(p^{s_{2}}-1)(p^{r-s_{2}}-\varepsilon p^{\frac{r}{2}-s_{2}})}{2(p^{s_{1}}-1)}$. Note that such a condition holds for the vectorial dual-bent functions listed in Section 2 (including bent functions belonging to $\mathcal{RF}$). The following corollary is directly from Theorem 3.

\begin{corollary}\label{Corollary2}
Let $F: V_{r} \rightarrow \mathbb{F}_{p^m}$ be a vectorial dual-bent function satisfying the Condition A and $F(cx)=c^{l}F(x), c\in \mathbb{F}_{p^m}^{*}, x \in V_{r}$ for some even positive integer $l$. Let $s_{1}, s_{2}$ be positive integers with $s_{2} \mid m, s_{1} \mid s_{2}$. Let $q=p^{s_{1}}$, $\lambda \in \mathbb{F}_{p^m}^{*}$ and $\widetilde{D}(F)=\{w^{[1]}, \dots, w^{[u]}\}$ be a subset of $D(F)=\{x \in V_{r}^{*}: Tr_{s_{2}}^{m}(\lambda F(x)) \text{ is a square in }\mathbb{F}_{p^{s_{2}}}^{*}\}$ (or $D(F)=\{x \in V_{r}^{*}: Tr_{s_{2}}^{m}(\lambda F(x)) \text{ is a non}$ $\text{-square in }\mathbb{F}_{p^{s_{2}}}^{*}\}$) such that (18) holds. Define
\begin{equation}\label{19}
  \mathcal{C}_{\widetilde{D}(F)}=\{c_{\alpha}=(\sum_{j=1}^{t}Tr_{s_{1}}^{r_{j}}(\alpha_{j}d_{j}))_{d=(d_{1}, \dots, d_{t}) \in \widetilde{D}(F)}: \alpha=(\alpha_{1}, \dots, \alpha_{t})\in V_{r}\}.
\end{equation}
Then $\mathcal{C}_{\widetilde{D}(F)}$ defined by (19) is a two-weight $q$-ary linear code with parameters $[\frac{(p^{s_{2}}-1)(p^{r-s_{2}}-\varepsilon p^{\frac{r}{2}-s_{2}})}{2(p^{s_{1}}-1)}, \frac{r}{s_{1}}]$ and the weight distribution is given in Table 5.
\end{corollary}
\begin{table}\label{5}
  \caption{The weight distribution of $\mathcal{C}_{\widetilde{D}(F)}$ defined by (19)}
  \begin{tabular}{|c|c|}
    \hline
    Hamming weight $a$ & Multiplicity $A_{a}$ \\
    \hline
    \hline
    $0$ & $1$ \\
    $\frac{ (p^{s_{2}}-1)p^{r-s_{1}-s_{2}}
    	\ +\varepsilon (\epsilon^{2s_2}-1) p^{\frac{r}{2}-s_1}}{2}$ & $\frac{p^{s_{2}}-1}{2}(p^{r-s_{2}}+\varepsilon p^{\frac{r}{2}-s_{2}})+p^{r-s_{2}}-1$ \\
    $\frac{(p^{s_{2}}-1)p^{r-s_{1}-s_{2}}
    	\ -\varepsilon(\epsilon^{2s_2}+1)p^{\frac{r}{2}-s_1}}{2}$ & $\frac{p^{s_{2}}-1}{2}(p^{r-s_{2}}-\varepsilon p^{\frac{r}{2}-s_{2}})$ \\
    \hline
  \end{tabular}
\end{table}

\begin{remark}
As we illustrate before, any bent function $f \in \mathcal{RF}$ satisfies the Condition A. When $s_{1}=s_{2}=1$ (which corresponds to $p$-ary linear code), the linear codes constructed in Theorem 3 and Corollary 2 reduce to the linear codes constructed in Theorem 19 and Corollary 28 of \cite{Tang}.
\end{remark}

When $q\neq p$, based on vectorial dual-bent functions listed by (1)-(6), we list the corresponding two-weight $q$-ary linear codes from Theorem 3 and Corollary 2.
\begin{itemize}
  \item Let $F: \mathbb{F}_{p^{r'}} \times \mathbb{F}_{p^{r'}}\rightarrow \mathbb{F}_{p^m}$ be defined as (1) (respectively, (2), (5)). Then $F$ is vectorial dual-bent satisfying the conditions in Theorem 3 and Corollary 2, and the corresponding value $\varepsilon=1$. Let $s_{1}, s_{2}$ be positive integers with $s_{1} \mid s_{2}$ and $s_{2} \mid m$, and let $\lambda \in \mathbb{F}_{p^m}^{*}$, $S=\{x^{2}: x \in \mathbb{F}_{p^{s_{2}}}^{*}\}$, $N=\mathbb{F}_{p^{s_{2}}}^{*} \backslash S$. Define
      \begin{equation*}
      \begin{split}
        &D(F)=\{x=(x_{1}, x_{2}) \in (\mathbb{F}_{p^{r'}}\times \mathbb{F}_{p^{r'}})^{*}: Tr_{s_{2}}^{r'}(\lambda a x_{1}x_{2}^{e}) \in S\}
        \end{split}
      \end{equation*}
      or
      \begin{equation*}
      \begin{split}
        &D(F)=\{x=(x_{1}, x_{2}) \in (\mathbb{F}_{p^{r'}}\times \mathbb{F}_{p^{r'}})^{*}: Tr_{s_{2}}^{r'}(\lambda a x_{1}x_{2}^{e}) \in N\},
        \end{split}
      \end{equation*}
      (respectively,
      \begin{equation*}
      \begin{split}
       & D(F)=\{x=(x_{1}, x_{2}) \in (\mathbb{F}_{p^{r'}}\times \mathbb{F}_{p^{r'}})^{*}: Tr_{s_{2}}^{r'}(\lambda a x_{1}L(x_{2}))\in S\}\\
      \end{split}
      \end{equation*}
      or \begin{equation*}
      \begin{split}
       & D(F)=\{x=(x_{1}, x_{2}) \in (\mathbb{F}_{p^{r'}}\times \mathbb{F}_{p^{r'}})^{*}: Tr_{s_{2}}^{r'}(\lambda a x_{1}L(x_{2}))\in N\},\\
      \end{split}
      \end{equation*}
      \begin{equation*}
      \begin{split}
       & D(F)=\{x=(x_{1}, x_{2})\in (\mathbb{F}_{p^{r'}} \times \mathbb{F}_{p^{r'}})^{*}: Tr_{s_{2}}^{m}(\lambda g(\gamma G(x_{1}x_{2}^{p^{r'}-2}))) \in S\}
      \end{split}
      \end{equation*}
      or
      \begin{equation*}
      \begin{split}
       & D(F)=\{x=(x_{1}, x_{2})\in (\mathbb{F}_{p^{r'}} \times \mathbb{F}_{p^{r'}})^{*}: Tr_{s_{2}}^{m}(\lambda g(\gamma G(x_{1}x_{2}^{p^{r'}-2}))) \in N\}
      \end{split}
      \end{equation*}
      ). Let $\widetilde{D}(F)=\{w^{[1]}, \dots, w^{[u]}\}$ such that (18) holds. Then the corresponding linear code $\mathcal{C}_{D(F)}$ defined by (17) is
  \begin{equation*}
  \begin{split}
  &\mathcal{C}_{D(F)}\\
  &=\{c_{\alpha}=(Tr_{s_{1}}^{r'}(\alpha_{1}d_{1}+\alpha_{2}d_{2}))_{d=(d_{1}, d_{2})\in D(F)}: \alpha=(\alpha_{1}, \alpha_{2}) \in \mathbb{F}_{p^{r'}}\times\mathbb{F}_{p^{r'}}\},
  \end{split}
  \end{equation*}
  which is a two-weight $p^{s_{1}}$-ary $[\frac{(p^{s_{2}}-1)}{2}(p^{2r'-s_{2}}- p^{r'-s_{2}}), \frac{2r'}{s_{1}}]$ linear code.
  The corresponding linear code $\mathcal{C}_{\widetilde{D}(F)}$ defined by (19) is a two-weight $p^{s_{1}}$-ary $[\frac{(p^{s_{2}}-1)(p^{2r'-s_{2}}-p^{r'-s_{2}})}{2(p^{s_{1}}-1)}, \frac{2r'}{s_{1}}]$ linear code.
  \item Let $F: \mathbb{F}_{p^{r}}\rightarrow \mathbb{F}_{p^m}$ be defined as (3) with $2m \mid r$. Then $F$ is vectorial dual-bent satisfying the conditions in Theorem 3 and Corollary 2, and the corresponding value $\varepsilon=-\epsilon^{{r}}\eta_{r}(a)$. Let $s_{1}, s_{2}$ be positive integers with $s_{1} \mid s_{2}$ and $s_{2} \mid m$, and let $\lambda \in \mathbb{F}_{p^m}^{*}$, $S=\{x^{2}: x \in \mathbb{F}_{p^{s_{2}}}^{*}\}$, $N=\mathbb{F}_{p^{s_{2}}}^{*} \backslash S$. Define $D(F)=\{x \in \mathbb{F}_{p^{r}}^{*}: Tr_{s_{2}}^{r}(\lambda ax^{2})\in S\}$ (or $D(F)=\{x \in \mathbb{F}_{p^{r}}^{*}: Tr_{s_{2}}^{r}(\lambda ax^{2})\in N\}$) and $\widetilde{D}(F)=\{w^{[1]}, \dots, w^{[u]}\}$ such that (18) holds. Then the corresponding linear code $\mathcal{C}_{D(F)}$ defined by (17) is
  \begin{equation*}
  \begin{split}
  &\mathcal{C}_{D(F)}=\{c_{\alpha}=(Tr_{s_{1}}^{r}(\alpha d))_{d\in D(F)}: \alpha \in\mathbb{F}_{p^{r}}\},
  \end{split}
  \end{equation*}
  which is a two-weight $p^{s_{1}}$-ary $[\frac{(p^{s_{2}}-1)}{2}(p^{r-s_{2}}+\epsilon^{{r}}\eta_{r}(a)p^{\frac{r}{2}-s_{2}}), \frac{r}{s_{1}}]$ linear code. The corresponding linear code $\mathcal{C}_{\widetilde{D}(F)}$ defined by (19) is a two-weight $p^{s_{1}}$-ary $[\frac{(p^{s_{2}}-1)(p^{r-s_{2}}+\epsilon^{{r}}\eta_{r}(a) p^{\frac{r}{2}-s_{2}})}{2(p^{s_{1}}-1)}, \frac{r}{s_{1}}]$ linear code.
  \item Let $F: \mathbb{F}_{p^m}^{t}\rightarrow \mathbb{F}_{p^m}$ be defined as (4) with $t$ even. Then $F$ is vectorial dual-bent satisfying the conditions in Theorem 3 and Corollary 2, and the corresponding value $\varepsilon=\epsilon^{mt}\eta_{m}(a_{1}\cdots a_{t})$. Let $s_{1}, s_{2}$ be positive integers with $s_{1} \mid s_{2}$ and $s_{2} \mid m$, and let $\lambda \in \mathbb{F}_{p^m}^{*}$, $S=\{x^{2}: x \in \mathbb{F}_{p^{s_{2}}}^{*}\}$, $N=\mathbb{F}_{p^{s_{2}}}^{*} \backslash S$. Define $D(F)=\{x=(x_{1}, \dots, x_{t}) \in (\mathbb{F}_{p^m}^{t})^{*}: \sum_{i=1}^{t}Tr_{s_{2}}^{m}(\lambda a_{i}x_{i}^{2}) \in S\}$ (or $D(F)=\{x=(x_{1}, \dots, x_{2t}) \in (\mathbb{F}_{p^m}^{t})^{*}: \sum_{i=1}^{t}Tr_{s_{2}}^{m}(\lambda a_{i}x_{i}^{2}) \in N\}$) and $\widetilde{D}(F)=\{w^{[1]}, \dots, w^{[u]}\}$ such that (18) holds. Then the corresponding linear code $\mathcal{C}_{D(F)}$ defined by (17) is
  \begin{equation*}
  \begin{split}
  &\mathcal{C}_{D(F)}=\{c_{\alpha}=(\sum_{i=1}^{t}Tr_{s_{1}}^{m}(\alpha_{j}d_{j}))_{d=(d_{1}, \dots, d_{t})\in D(F)}: \alpha=(\alpha_{1}, \dots, \alpha_{t}) \in\mathbb{F}_{p^m}^{t}\},
  \end{split}
  \end{equation*}
  which is a two-weight $p^{s_{1}}$-ary $[\frac{(p^{s_{2}}-1)}{2}(p^{mt-s_{2}}-\epsilon^{mt}\eta_{m}(a_{1}\cdots a_{t})p^{\frac{mt}{2}-s_{2}}), \frac{mt}{s_{1}}]$ linear code. The corresponding linear code $\mathcal{C}_{\widetilde{D}(F)}$ defined by (19) is a two-weight $p^{s_{1}}$-ary $[\frac{(p^{s_{2}}-1)(p^{mt-s_{2}}-\epsilon^{mt}\eta_{m}(a_{1}\cdots a_{t}) p^{\frac{mt}{2}-s_{2}})}{2(p^{s_{1}}-1)}, \frac{mt}{s_{1}}]$ linear code.
  \item Let $F: \mathbb{F}_{p^{r'}} \times \mathbb{F}_{p^{r''}} \times \mathbb{F}_{p^{r''}}\rightarrow \mathbb{F}_{p^m}$ be defined as (6) for which $2m \mid r'$ and $\alpha_{j}, 1\leq j\leq 3$ are all squares or all non-squares in $\mathbb{F}_{p^{r'}}^{*}$. Then $F$ is vectorial dual-bent satisfying the conditions in Theorem 3 and Corollary 2, and the corresponding value $\varepsilon=-\epsilon^{r'}\eta_{r'}(\alpha_{1})$. Let $s_{1}, s_{2}$ be positive integers with $s_{1} \mid s_{2}$ and $s_{2}\mid m$, $\lambda \in \mathbb{F}_{p^m}^{*}$, $S=\{x^{2}: x \in \mathbb{F}_{p^{s_{2}}}^{*}\}$, $N=\mathbb{F}_{p^{s_{2}}}^{*} \backslash S$. Define $D(F)=\{x=(x_{1}, x_{2}, x_{3}) \in (\mathbb{F}_{p^{r'}}\times \mathbb{F}_{p^{r''}}\times \mathbb{F}_{p^{r''}})^{*}: Tr_{s_{2}}^{m}(\lambda F(x_{1}, x_{2}, x_{3}))\in S\}$ (or $D(F)=\{x=(x_{1}, x_{2}, x_{3}) \in (\mathbb{F}_{p^{r'}}\times \mathbb{F}_{p^{r''}}\times \mathbb{F}_{p^{r''}})^{*}: Tr_{s_{2}}^{m}(\lambda F(x_{1}, x_{2}, x_{3})) \in N\}$) and $\widetilde{D}(F)=\{w^{[1]}, \dots, w^{[u]}\}$ such that (18) holds. Then the corresponding linear code $\mathcal{C}_{D(F)}$ defined by (17) is
  \begin{equation*}
  \begin{split}
  &\mathcal{C}_{D(F)}=\{c_{\alpha}=(Tr_{s_{1}}^{r'}(\alpha_{1}d_{1})+Tr_{s_{1}}^{r''}(\alpha_{2}d_{2}+\alpha_{3}d_{3}))_{d=(d_{1}, d_{2}, d_{3}) \in D(F)}: \\
  &\ \ \ \ \ \ \ \ \ \ \ \ \ \alpha=(\alpha_{1}, \alpha_{2}, \alpha_{3}) \in \mathbb{F}_{p^{r'}}\times \mathbb{F}_{p^{r''}}\times \mathbb{F}_{p^{r''}}\},
  \end{split}
  \end{equation*}
  which is a two-weight $p^{s_{1}}$-ary $[\frac{(p^{s_{2}}-1)}{2}(p^{r'+2r''-s_{2}}+\epsilon^{r'}\eta_{r'}(\alpha_{1})p^{\frac{r'}{2}+r''-s_{2}}),$ $\frac{r'+2r''}{s_{1}}]$ linear code.
  The corresponding linear code $\mathcal{C}_{\widetilde{D}(F)}$ defined by (19) is a two-weight $p^{s_{1}}$-ary $[\frac{(p^{s_{2}}-1)(p^{r'+2r''-s_{2}}+\epsilon^{r'}\eta_{r'}(\alpha_{1}) p^{\frac{r'}{2}+r''-s_{2}})}{2(p^{s_{1}}-1)}, \frac{r'+2r''}{s_{1}}]$ linear code.
\end{itemize}

We give two examples.
\begin{example} \label{Example5}
Let $p=3, s_{1}=s_{2}=2, m=4, r=8$ and $\lambda=1$, $F: \mathbb{F}_{3^4}\times \mathbb{F}_{3^4}\rightarrow \mathbb{F}_{3^4}$ be defined as $F(x_{1}, x_{2})=x_{1}x_{2}^{7}$. The linear code  $\mathcal{C}_{\widetilde{D}(F)}$ defined by (19) is
\begin{equation*}
  \mathcal{C}_{\widetilde{D}(F)}=\{c_{\alpha}=(Tr_{2}^{4}(\alpha_{1}d_{1}+\alpha_{2}d_{2}))_{d=(d_{1}, d_{2}) \in \widetilde{D}(F)}: \alpha=(\alpha_{1}, \alpha_{2}) \in \mathbb{F}_{3^4} \times \mathbb{F}_{3^4}\},
\end{equation*}
where $\widetilde{D}(F)=\{w^{[1]}, \dots, w^{[360]}\}$ such that $D(F)=\bigcup_{i=1}^{360}w^{[i]}\mathbb{F}_{9}^{*}$, where $D(F)=\{x=(x_{1}, x_{2}) \in (\mathbb{F}_{3^4} \times \mathbb{F}_{3^4})^{*}: Tr_{2}^{4}(x_{1}x_{2}^{7}) \ \text{is a square in } \mathbb{F}_{9}^{*}\}$ (or $D(F)=\{x=(x_{1}, x_{2}) \in (\mathbb{F}_{3^4} \times \mathbb{F}_{3^4})^{*}: Tr_{2}^{4}(x_{1}x_{2}^{7}) \ \text{is a non-square in } \mathbb{F}_{9}^{*}\}$). Then $\mathcal{C}_{\widetilde{D}(F)}$ is a two-weight $9$-ary linear code with parameters $[360, 4, 315]$ and the weight enumerator is $1+2880z^{315}+3680z^{324}$.
\end{example}

\begin{example}\label{Example6}
Let $p=11, s_{1}=s_{2}=m=2, r=4$ and $\lambda=1$, $F: \mathbb{F}_{11^2} \times \mathbb{F}_{11^2}\rightarrow \mathbb{F}_{11^2}$ be defined as
$F(x_{1}, x_{2})=g(x_{1}x_{2}^{119})$, where $g: \mathbb{F}_{11^2}\rightarrow \mathbb{F}_{11^2}$ is a permutation defined by $g(x)=x^{61}+2x$. The linear code  $\mathcal{C}_{\widetilde{D}(F)}$ defined by (19) is
\begin{equation*}
  \mathcal{C}_{\widetilde{D}(F)}=\{c_{\alpha}=(\alpha_{1}d_{1}+\alpha_{2}d_{2})_{d=(d_{1}, d_{2}) \in \widetilde{D}(F)}: \alpha=(\alpha_{1}, \alpha_{2}) \in \mathbb{F}_{11^2} \times \mathbb{F}_{11^2}\},
\end{equation*}
where $\widetilde{D}(F)=\{w^{[1]}, \dots, w^{[60]}\}$ such that $D(F)=\bigcup_{i=1}^{60}w^{[i]}\mathbb{F}_{11^2}^{*}$, where $D(F)$ $=\{x=(x_{1}, x_{2}) \in (\mathbb{F}_{11^2} \times \mathbb{F}_{11^2})^{*}:  g(x_{1}x_{2}^{119}) \ \text{is a square in } \mathbb{F}_{11^2}^{*}\}$ (or $D(F)=\{x=(x_{1}, x_{2}) \in (\mathbb{F}_{11^2} \times \mathbb{F}_{11^2})^{*}: g(x_{1}x_{2}^{119})\ \text{is a non-square in } \mathbb{F}_{11^2}^{*}\}$). Then $\mathcal{C}_{\widetilde{D}(F)}$ is a two-weight $121$-ary MDS linear code with parameters $[60, 2, 59]$, which is optimal and the weight enumerator is $1+7200z^{59}+7440z^{60}$.
\end{example}

Based on Lemmas 2 and 4, we have the following proposition. For the convenience of expression, we both use $\mathcal{C}$ to represent the linear codes constructed in Theorem 3 and Corollary 2.

\begin{proposition}\label{Proposition5}
Let $\mathcal{C}$ be the $p^{s_{1}}$-ary linear code constructed in Theorem 3 or Corollary 2 with $s_{2}<\frac{r}{2}$ or $s_{2}\neq s_{1}$, and $s_{2}\neq 1$ if $p=3, r=4, \varepsilon=1$. Let $G=[g_{0}, g_{1}, \dots, g_{n-1}]$ be a generator matrix of $\mathcal{C}$. In the secret sharing scheme based on $\mathcal{C}^{\bot}$, there are altogether $p^{r-s_{1}}$ minimal access sets.
\begin{itemize}
  \item If $\mathcal{C}$ is constructed in Theorem 3:
     For any $1\leq i \leq n-1$, if $g_{i}$ is a multiple of $g_{0}$, then the participant $P_{i}$ must be in every minimal access set. If $g_{i}$ is not a multiple of $g_{0}$, then the participant $P_{i}$ must be in $(p^{s_{1}}-1)p^{r-2s_{1}}$ out of $p^{r-s_{1}}$ minimal access sets.
  \item If $\mathcal{C}$ is constructed in Corollary 2:
     For any fixed $1\leq l \leq min\{\frac{r}{s_{1}}-1, d^{\bot}-2\}$, every set of $l$ participants is involved in $p^{r-s_{1}(l+1)}(p^{s_{1}}-1)^{l}$ out of $p^{r-s_{1}}$ minimal access sets, where $d^{\bot}$ denotes the minimum distance of $\mathcal{C}^{\bot}$.
\end{itemize}
\end{proposition}
\textbf{Proof.}
From Lemma 2, it is easy to verify that the constructed $p^{s_{1}}$-ary linear code of Theorem 3 or Corollary 2 is minimal when $s_{2}<\frac{r}{2}$ or $s_{2}\neq s_{1}$, and $s_{2}\neq 1$ if $p=3, r=4, \varepsilon=1$. If $\mathcal{C}$ is constructed in Theorem 3, then obviously $d^{\bot} \neq 1$ and from $F(x)=F(-x)$ we have $-D(F)=D(F)$, which implies that $d^{\bot}=2$. If $\mathcal{C}$ is constructed in Corollary 2, then obviously $d^{\bot} \neq 1$ and if $d^{\bot}=2$, then there exist $d=(d_{1}, \dots, d_{t}), d'=(d'_{1}, \dots, d'_{t})\in \widetilde{D}(F)$ and $z \in \mathbb{F}_{p^{s_{1}}}^{*}$ such that $\sum_{j=1}^{t}Tr_{s_{1}}^{r_{j}}(\alpha_{j}d_{j})+z\sum_{j=1}^{t}Tr_{s_{1}}^{r_{j}}(\alpha_{j}d'_{j})=0$ for all $\alpha=(\alpha_{1}, \dots, \alpha_{t})\in V_{r}$, which implies that $d+zd'=0$ and this is impossible by the definition of $\widetilde{D}(F)$ given by (18). Therefore, $d^{\bot}\geq 3$ if $\mathcal{C}$ is constructed in Corollary 2. Then the conclusion follows from Lemma 4. \ $\Box$

\section{Conclusion}
In this paper, by using vectorial dual-bent functions, we provide several constructions of $q$-ary linear codes with two or three weights whose weight distributions are determined (Theorems 1, 2 and 3 and Corollaries 1 and 2), where $q$ is a power of an odd prime $p$. We illustrate that a class of three-weight linear codes in \cite{Carlet2}, two classes of two-weight linear codes in \cite{Tang} and two classes of two-weight linear codes in \cite{Tang2} can be obtained by our constructions. In some special cases, our constructed linear codes are optimal. In addition, we obtain secret sharing schemes with interesting access structures based on the constructed linear codes (Propositions 3, 4 and 5).

\section*{Acknowledgements}
This research is supported by the National Key Research and Development Program of China (Grant No. 2018YFA0704703), the National Natural Science Foundation of China (Grant Nos. 12141108 and 61971243), the Natural Science Foundation of Tianjin (20JCZDJC00610), the Fundamental Research Funds for the Central Universities of China (Nankai University), and the Nankai Zhide Foundation, the Natural Science Foundation of Hebei Province (Grant No. A2021205027), the Science Foundation of Hebei Normal University (Grant No. L2021B04).






\end{document}